\documentclass[useAMS,usenatbib]{mn2e}
\usepackage{graphicx}
\pdfoutput=1

\title[Diffuse continuum transfer in H{\sc\,ii} regions]{Diffuse
continuum transfer in H{\sc\,ii} regions\thanks{(c) Crown Copyright 2009/MoD}}

\author[R. J. R. Williams and W. J. Henney]{R. J. R. Williams$^1$ and
W. J. Henney$^2$\thanks{E-mail: robin.williams@awe.co.uk (RJRW);
w.henney@astrosmo.unam.mx (WJH)}\\
$^1$AWE Aldermaston, Reading, Berkshire, RG7 4PR\\
$^2$Centro de Radioastronom\'{\i}a y Astrof\'{\i}sica, Universidad
Nacional Aut\'{o}noma de M\'{e}xico, Morelia, M\'{e}xico}
\begin{document}
\date{Accepted xxx. Received xxx; in original form xxx}
\pagerange{\pageref{firstpage}--\pageref{lastpage}} \pubyear{2009}
\maketitle
\label{firstpage}
\begin{abstract}
We compare the accuracy of various methods for determining the
transfer of the diffuse Lyman continuum in H{\sc\,ii} regions, by
comparing them with a high-resolution discrete-ordinate integration.
We use these results to suggest how, in multidimensional dynamical
simulations, the diffuse field may be treated with acceptable accuracy
without requiring detailed transport solutions.

The angular distribution of the diffuse field derived from the
numerical integration provides insight into the likely effects of the
diffuse field for various material distributions.
\end{abstract}


\section{Introduction}

In this paper, we model the diffuse field structure of astrophysical
H{\sc\,ii} regions.  Recombinations within these regions emit
radiation in the various hydrogen line spectra and Balmer and higher
continua which is an observed characteristic of them.  Emission in the
Lyman continuum, however, is energetic enough to ionize other hydrogen
atoms, and so is trapped within the nebula.  This radiation field is
believed to have a significant fraction of the energy density of the
direct ionizing continuum in some parts of the H{\sc\,ii} region, and
so it is important to model it correctly.  This is particularly
relevant when modelling the complex dynamics of the nebulae, or the
emission from the internal features such as the tails of cometary
globules in the Helix planetary nebula \citep{odell07}.

\cite{ritze05} has suggested that diffuse fields may be particularly
important at the edges of H{\sc\,ii} regions, in some regimes.  This
would seem to suggest that the diffuse field may have a more
significant impact on their overall evolution than previously assumed.
However, Ritzerveld uses a simple outward-only treatment of the
diffuse radiation transport, and also assumes that the absorption
coefficient is comparable for diffuse and direct radiation fields,
although it is acknowledged that in reality the direct photons are
likely to have a harder spectrum and will thus be significantly more
penetrating.

To investigate the validity of these conclusions, in this paper we
apply detailed discrete-ordinate angular integration to investigate
the validity of several approximate numerical transport schemes.  To
simplify the problem, we assume a pure-hydrogen nebula, without dust,
and use a simple two-frequency approximation to the radiation flow.

With our more detailed modelling, we find that the diffuse field can
indeed dominate for the situations Ritzerveld describes.  However,
where the diffuse field dominates, it will typically also be
outwardly-beamed and therefore be indistinguishable from the direct
field for most purposes.  For most astrophysically relevant
conditions, the usual on-the-spot approximation is shown to give accurate
results through most of a spherical nebula, except for a region close
to the star where it {\it overestimates}\/ the diffuse field.

We also assess the accuracy of the Eddington diffusion approximation
for the diffuse field transfer, which may be a useful means of
modelling diffuse transport effects in multidimensional simulations.

In the present paper, we neglect the effects of dust and heavy
elements in the H{\sc\,ii} region.  This is a reasonable assumption
for the case of cosmological H{\sc\,ii} regions; however, there is
observational evidence for the importance of dust absorption within
the ionized gas in galactic H{\sc\,ii} regions
\citep{cesar00,robbe05}.  We briefly discuss the impact dust might
have on our results.

\section{Governing equations and physical conditions}

We use a simple model of H{\sc\,ii} regions to study the diffuse field
structure.  We consider steady, spherically symmetric pure-hydrogen
regions, with the radiation field in two frequency components: higher
frequency radiation propagating directly from a central point source,
and lower frequency diffuse radiation.  We assume that the material
within the H{\sc\,ii} region is maintained at a constant temperature
in ionization equilibrium.

In the absence of dust, the governing equations are then
\begin{eqnarray}
{dL_{\rm dir}/dr} &=& -n(1-x)a_0L_{\rm dir} \label{e:dir} \\
{dI_{\rm dif}/ds} &=& -n(1-x)a_1I_{\rm dif} + {\alpha_1\over 4\pi} n^2 x^2 
\label{e:dif}\\
\alpha_{\rm A} n^2 x^2 &=& n(1-x)\left({a_0L_{\rm dir}\over 4\pi r^2}
+ a_1\int I_{\rm dif}(r,\Omega'){\,d\Omega'}\right),\label{e:ion}
\end{eqnarray}
corresponding to the transfer of direct radiation from a point source,
$L_{\rm dir}$, the diffuse radiation field $I_{\rm dif}(r,\theta)$,
and balance of ionization and recombination for ionization fraction
$x$.  Here $L$ is the specific luminosity and $I$ the radiation
intensity specified in terms of the {\it number}\/ of photons, rather
than the radiation energy, as this is more natural in the present
application.

Integrating equations~(\ref{e:dir})--(\ref{e:ion}) over volume, and
angle where appropriate, and applying the boundary conditions $L_{\rm
dir} = L_\star$ at $r=0$, $I_{\rm dif} = 0$ as $r\to\infty$, we find
that
\begin{equation}
L_\star = \int \alpha_{\rm B} n^2 x^2 dV,
\end{equation}
where $\alpha_{\rm B} = \alpha_{\rm A} -\alpha_1$.  This has a simple
interpretation, in that the number of ionizing photons into the nebula
balances the number of recombinations which destroy such photons
(rather than re-emitting a Lyman continuum photon).

If the ionization front is thin, as is typically the case, then a
characteristic Str\"omgren radius for a spherical H{\sc\,ii} region,
$r_{\rm S}$, is given by
\begin{equation}
L_\star = \int_0^{r_{\rm S}} \alpha_{\rm B} n^2 4\pi r^2 dr,\label{e:rstrom}
\end{equation}
independent of diffuse radiation transport effects.

From \cite{agn3}, we take values of the recombination coefficients at
$10^4{\rm\,K}$ of $\alpha_{\rm A} =
4.18\times10^{-13}{\rm\,cm^3\,s^{-1}}$ and $\alpha_{\rm 1} =
1.58\times10^{-13}{\rm\,cm^3\,s^{-1}}$.  We assume that the ionization
cross section for the diffuse radiation is the Lyman limit cross
section $a_1 = 6\times10^{-18}{\rm\,cm^2}$, while for the direct
photons, we take either $a_0 = 1\times10^{-18}{\rm\,cm^2}$ (which
allows for the largest likely effect of the higher average frequency
of the direct photons) or $a_0 = 6\times10^{-18}{\rm\,cm^2}$
\citep[corresponding to the case which was primarily considered
by][]{ritze05}.  In practice, the effective absorption coefficients
will depend on the spectrum of the star and the recombination
continuum, and their modification due to intervening absorption
processes: the uncertainty is enhanced by the sharp dependence of the
absorption cross section on the photon frequency above threshold
\citep[see Figure 2.2 in][]{agn3}.

The calculations are normalized to a star emitting ionizing photons at
a rate $10^{48}{\rm\,s^{-1}}$, generating an equilibrium Str\"omgren
sphere of radius $r_{\rm S}=1{\rm\,pc}$ in all cases.  This
corresponds to a density of $n_{\rm H} = n_{\rm e} =
180{\rm\,cm^{-3}}$ for the uniform density case.

For direct comparison to Ritzerveld's work, we consider three main
density distributions: uniform, and proportional to $r^{-1}$ or
$r^{-2}$ beyond an empty core of radius $10^{-6}\times r_{\rm S}$ and
$0.05\times r_{\rm S}$, respectively.

\section{Solution methods}

\subsection{OTS and the diffusion limit}

The simplest approximation widely used in astrophysics is the case B
approximation.  Here, the diffuse field is assumed to be absorbed at
the same point as it is emitted, so equation~(\ref{e:dif}) is replaced
by the on-the-spot (OTS) approximation
\begin{equation}
n(1-x)a_1I_{\rm dif} = {\alpha_1\over4\pi} n^2 x^2.\label{e:ots}
\end{equation}
Note that the diffuse radiation intensity is isotropic in this
approximation.  With this assumption, equation~(\ref{e:ion})
simplifies to
\begin{equation}
\alpha_{\rm B} n^2 x^2 = {n(1-x)a_0L_{\rm dir}\over 4\pi r^2},
\label{e:casebbalance}
\end{equation}
and so
\begin{equation}
{dL_{\rm dir}/dr} = -4\pi r^2\alpha_{\rm B} n^2 x^2.\label{e:ldirots}
\end{equation}
Since the hydrogen ionization fraction is close to unity through most
of an astrophysical H{\sc\,ii} region, this equation may then be
integrated to define the Str\"omgren radius, in agreement with
equation~(\ref{e:rstrom}).  The ratio of the diffuse field photon
density to that of the ionizing field is
\begin{equation}
f_{\rm dif, OTS} = {\int I_{\rm dif}d\Omega\over L_{\rm dir}/4\pi r^2}
= {\alpha_1 a_0\over \alpha_B a_1},
\end{equation}
which is roughly $60$ per cent for equal absorption cross sections, or
$\sim 10$ per cent if we assume a lower absorption cross section for
direct photons.  The ratio of the diffuse flux through a surface (of
any orientation) to the direct flux through a surface directed towards
the star will be one quarter of this value \citep[as in the analysis
of][who assume $a_0=a_1$, and thus find $F_{\rm D} =
0.15F_\star$]{canto98}.

If we wish to extend this analysis, we need to obtain a better
estimate for $J_{\rm dif} = \int I_{\rm dif} d\Omega$ to insert in
equation~(\ref{e:ion}).  Writing equation~(\ref{e:dif}) as
\begin{equation}
{dI_{\rm dif}\over ds} = -\kappa I_{\rm dif} + {\epsilon\over 4\pi},
\label{e:gentran}
\end{equation}
where $\kappa=n(1-x)a_1$ is the opacity and $\epsilon=\alpha_1n^2 x^2$
is the volume emissivity, then we can write
\begin{equation}
J_{\rm dif} = {\epsilon\over\kappa} - {1\over\kappa}\int d\Omega 
{dI_{\rm dif}\over ds}
\end{equation}
This exact expression remains dependent on $I_{\rm dif}$.  If we apply
the on-the-spot approximation $I_{\rm dif}\simeq
\epsilon/4\pi\kappa$ to truncate the hierarchy at this
approximation, the integration over angle is zero, which shows that
the case B approximation is accurate to second order.  

To obtain more accurate results, however, we can substitute for
$I_{\rm dif}$ again using equation~(\ref{e:gentran}).  This gives
\begin{equation}
J_{\rm dif} = {\epsilon\over\kappa} + {1\over\kappa}\int {d\over ds}
\left({1\over \kappa} {dI_{\rm dif}\over ds} - {\epsilon\over
4\pi\kappa}\right)d\Omega.
\end{equation}
Again, the second term in the integral will be zero.  Assuming the
remaining $I_{\rm dif}$ term is also isotropic to first order, it will
be a function of radius alone, so
\begin{eqnarray}
{dI_{\rm dif}\over ds} &=& \cos\theta {dI_{\rm dif}\over dr}\\
{d\over ds}\left({1\over\kappa}{dI_{\rm dif}\over ds}\right) &=&
	\cos^2\theta {d\over dr}\left({1\over\kappa}
	{dI_{\rm dif}\over dr}\right)
+	\sin^2\theta {1\over\kappa r}{dI_{\rm dif}\over dr}
\end{eqnarray}
and so
\begin{equation}
J_{\rm dif} \simeq {\epsilon\over\kappa} + {1\over \kappa r^2} {d\over
dr} {r^2\over 3\kappa} {dJ_{\rm dif}\over dr},
\end{equation}
corresponding to Eddington approximation $J=3K$
\citep[e.g.\@][]{mihal99}.  Thus given a distribution of ionization,
we can solve for the diffuse field using this expression, in which the
diffusive transport corrects the OTS intensity.  Note, that since
$x\sim 1$ in the centre of the H{\sc\,ii} region, the free path
$1/\kappa$ may be large.

\subsection{Ritzerveld's outward-only approximation}

\cite{ritze05} solves an outward-only system
\begin{eqnarray}
{dL_{\rm dir}\over dr} &=& -\xi\alpha_{\rm A} n^2 r^{d-1}
	{cL_{\rm dir}\over cL_{\rm dir}+L_{\rm dif}}\label{e:ritzerdir}\\
{dL_{\rm dif}\over dr} &=& -\xi\alpha_{\rm A} n^2 r^{d-1} 
	{L_{\rm dif}\over cL_{\rm dir}+L_{\rm dif}} 
	+ \xi r^{d-1} \alpha_1 n^2,\label{e:ritzerdif}
\end{eqnarray}
where for slab symmetry $d=1$, $\xi=1$, for cylindrical symmetry
$d=2$, $\xi = 2\pi$ and for spherical symmetry $d=3$, $\xi = 4\pi$.
Note that these equations are given here for the total ionization
intensity integrated over the surface at the specified radius.
Ritzerveld states that they apply to the density of photons per unit
volume, but this is clearly not the case, as if all the recombination
coefficients are zero, then $L_{\rm dir} = L_{\rm dif} = {\it const}$.
This system approximates the transport in the limit of nearly complete
ionization within the H{\sc\,ii} region.  The factor $c=a_0/a_1$
allows for the preferential absorption of the diffuse photons.

The results presented by \cite{ritze05} can be generalized to the case
of arbitrary ratios of absorption cross section.  Adding equations
(\ref{e:ritzerdir}) and (\ref{e:ritzerdif}), we find
\begin{equation}
{d\over dr}\left(L_{\rm dir}+L_{\rm dif}\right) =
	 -\xi\alpha_{\rm B} n^2 r^{d-1},\label{e:ritzersum}
\end{equation}
and hence
\begin{equation}
L_{\rm tot} \equiv L_{\rm dir}+L_{\rm dif} =
	L_\star-\int^r \xi\alpha_{\rm B} n^2 r^{d-1} dr. \label{e:itot}
\end{equation}

A general (implicit) expression for the direct and diffuse fields can
be derived in terms of the total field, independent of the density
distribution or geometry.  Dividing equation~(\ref{e:ritzersum}) by
equation~(\ref{e:ritzerdir}), we find
\begin{equation}
{dL_{\rm tot}\over dL_{\rm dir}} = 
{\alpha_{\rm B}\over c\alpha_{\rm A}}
\left({L_{\rm tot}\over L_{\rm dir}} + c - 1\right),
\end{equation}
and hence
\begin{equation}
L_{\rm tot} = {1\over\alpha_{\rm A}c-\alpha_{\rm B}}
\left[ \alpha_1 c L_\star 
\left(L_{\rm dir}\over L_\star\right)^{\alpha_{\rm B}/c\alpha_{\rm A}}
-(1-c)\alpha_{\rm B}L_{\rm dir}\right].\label{e:gendirect}
\end{equation}
Given $L_{\rm tot}(r)$ from equation~(\ref{e:itot}) above,
equation~(\ref{e:gendirect}) may then be solved for $L_{\rm dir}$
(explicitly in the case $c=1$, to recover the analytic solutions which
Ritzerveld presents), and then $L_{\rm dif} = L_{\rm tot}-L_{\rm
dir}$.

\subsection{Full transfer}

We use an iterative technique to find solutions, using similar
techniques to those applied elsewhere \citep{humme63,rubin68}.  Here,
we integrate the diffuse field using a large number of coaxial zones.

We first derive an approximate ionization structure using the case B
approximation, and use this to estimate the outward-going diffuse
field components.  We integrate the diffuse field transport inwards
through this structure.  As part of this sweep, we calculate an
improved estimate for the ionization structure taking account of the
new inward radiation field beams.  We then integrate the direct
radiation and outward-going diffuse field outward through the
resulting structure, using the inward diffuse field at its closest
point to the star as the initial condition for the outward beams,
updating the ionization structure again.  We iterate these inward and
outward sweeps until the ionization structure has converged -- this
convergence is rapid in practice.

The results for the ionization structure are very sensitive to the
details of the numerical scheme for the case where the density varies
as $r^{-2}$, in particular the Str\"omgren radius can vary
significantly.  As noted by \cite{franc90}, the ionization front can
in this case escape to infinity: having the ionization front at 20
times the inner radius of the density distribution requires a rather
precisely tuned value of the incident radiation field.  This feeds
through to the numerical sensitivity.

As a result of this, considerable care has to be taken in the
radiation field integration to ensure the rate at which photons are
removed from the radiation field is consistent with the ionization
rate within the zone \citep{abele99,willi02,melle06,whale06}.  To do
this, we will adapt the approach described by \cite{willi02} to the
case of multiple incident beams of radiation and diffuse sources.

We specify the radiation field in coaxial cylindrical zones, which
intersect spherical shells within which the physical variables are
assumed to be uniform.  We take the $i$-th component of the density
field to occupy a spherical shell of outer radius $r_i = i\Delta$, and
the $j$-th group of diffuse photons to be those with an impact
parameter between $R_{j-1}$ and $R_j$, where $R_j = j\Delta$.  

Integrating equation~(\ref{e:ion}) over a spatial zone, we have
\begin{eqnarray}
\alpha_{\rm A} n^2 x^2 \Delta V_i
&=& \int_{r_{i-1}}^{r_i} n(1-x)\left({a_0L_{\rm dir}\over 4\pi r^2}\right.+
\nonumber\\
&&\quad \left.a_1\int I_{\rm dif}(r,\Omega'){\,d\Omega'}\right)\,dV,
\end{eqnarray}
where
\begin{equation}
\Delta V_i = {4\pi\over 3}\left(r_i^3- r_{i-1}^3\right).
\end{equation}
Substituting from the transport equations, (\ref{e:dir}) and~(\ref{e:dif}),
then gives
\begin{eqnarray}
\alpha_{\rm B} n^2 x^2 \Delta V_i
&=&-\left[L_{\rm dir}\right]_{r_{i-1}}^{r_i}\nonumber\\
&&\quad
-\int_{r_{i-1}}^{r_i} \int {dI_{\rm dif}\over ds} \,d\Omega' 4\pi r^2 \,dr,
\label{e:disceqm}
\end{eqnarray}
where the problem is uniform in space, so the angular integral over
the zone volume is trivial.  Combining the spatial integral $r^2 dr$
with the beam solid angle integral $d\Omega'$, however, yields an
integral which may be taken as over the zone volume for beams which
are incident from a single direction.  This may be discretized as the
difference in diffuse flux within each radiation bin across the zone,
\begin{equation}
\int_{r_{i-1}}^{r_i} \int {dI_{\rm dif} \over ds} \,d\Omega' 4\pi r^2 \,dr
= \int_{\Omega_i} {dI_{\rm dif}\over ds}(b,s) 4\pi \cdot 2\pi b \,db\,ds,
\end{equation}
where the integral may now be approximated as the change in the
diffuse flux within the cylinder
\begin{eqnarray}
I_{{\rm dif},j} \Delta A_j &=& \int_{b_{j-1}}^{b_j} I_{\rm dif} 2\pi b\,db\\ 
\int_{s_{i-1}}^{s_i} {dI_{\rm dif} \over ds} ds &=& I_{\rm
dif,\it i} - I_{\rm dif,\it i-1}
\end{eqnarray}
as the photons pass across the radial zone.  Again, this has the
interpretation that the difference between ionizing photons into and
out of the zone is balanced with the rate of recombinations in the
zone which destroy such photons.

In order to evaluate the differences in the direct and diffuse fluxes
on either side of the zone, we discretize the transport equations.
There are a range of choices for this discretization, which give
different forms for the equations to be solved to evaluate the
ionization fraction \citep{willi02}.  

As the source function from the recombinations means that there is no
simple closed form solution to the ionization equilibrium problem in
the zone, we instead choose a scheme where the outgoing beams are
guaranteed to be positive.  We integrate the transfer
equation~(\ref{e:dif}) across a uniform zone to give
\begin{eqnarray}
I(s+\Delta s) &\simeq& I(s) \exp\left[-n(1-x)a_1\Delta s\right]+\nonumber\\
&&\quad
S_{\rm dif}(x)\left(1-\exp\left[-na_1(1-x)\Delta s\right]\right),
\end{eqnarray}
where
\begin{equation}
S_{\rm dif}(x) = {\alpha_1 n x^2\over4\pi a_1(1-x)}\label{e:sdif}
\end{equation}
is the diffuse field source function.  The change in ionizing flux
across the zone required for the ionization balance equation in the
form~(\ref{e:disceqm}) is then
\begin{eqnarray}
I(s+\Delta s)-I(s) &=& \left[S_{\rm dif}(x)-I(s)\right]\times\nonumber\\
&&\quad\left(1-\exp\left[-n(1-x)a_1\Delta s\right]\right).
\end{eqnarray}
This expression must then be integrated over the cylindrical zone of
impact parameters and the solid angle of photon directions to obtain a
finite difference in the form needed to be used in
equation~(\ref{e:disceqm}).  On the right hand side, we make the
first-order assumption that a single mean path length applies to all
photons within the photon zone as they cross the radial zone.

In order to perform the discrete integrals, we need to determine the
volume of intersection between the radiation and spatial zones.  The
volume between these cylindrical and spherical shells, each of
thickness $\Delta$, is contained in two disjoint components, apart
from the case $i=j$.  The radiation field in the latter case can still
be advanced by the same formalism by splitting the zone in two at the
intersection of the radiation beam with the equatorial plane.  Note,
however, that the additional radiation flux values introduced into the
diffuse field arrays as a result must not be included when calculating
radiation field moments at the radial zone interfaces.  

Each component has volume
\begin{equation}
v_{i,j} = {1\over 2}
\left(V_{i-1,j}+V_{i,j-1}-V_{i,j}-V_{i-1,j-1}\right),
\end{equation}
where $V_{i,j}$ is the volume within a coaxial sphere of radius $r_i$
but outside a cylinder of radius $R_j < r_i$,
\begin{equation}
V_{i,j} = {4\pi\over 3}\left(r_i^2-R_j^2\right)^{3/2}.
\end{equation}
The area of the cylindrical shell is
\begin{equation}
a_j = \pi (R_j^2 - R_{j-1}^2) = (2j-1)\pi \Delta^2.
\end{equation}
Hence the average path length of photons within the cylindrical shell
$j$ through one of the the intersections with the spherical shell is
\begin{equation}
\Delta s_{i,j} = {v_{i,j}\over a_j}.
\end{equation}
We use these distances as the chord lengths for propagation of the
diffuse radiation propagation in the cylindrical shell labelled $j$
through its intersection with the spherical shell labelled $i$ in the
ionization equilibrium solution, in order to obtain volume
consistency.  The resulting expression for the zone ionization,
including all direct and diffuse beams, is solved using Newton-Raphson
iteration.

In all cases, we verify that at convergence the number of case B
recombinations within the modelled H{\sc\,ii} region balances the
incoming ionizing flux to high accuracy.

\section{Results}

\begin{figure*}
\begin{tabular}{cc}
\includegraphics[width=6cm,angle=270]{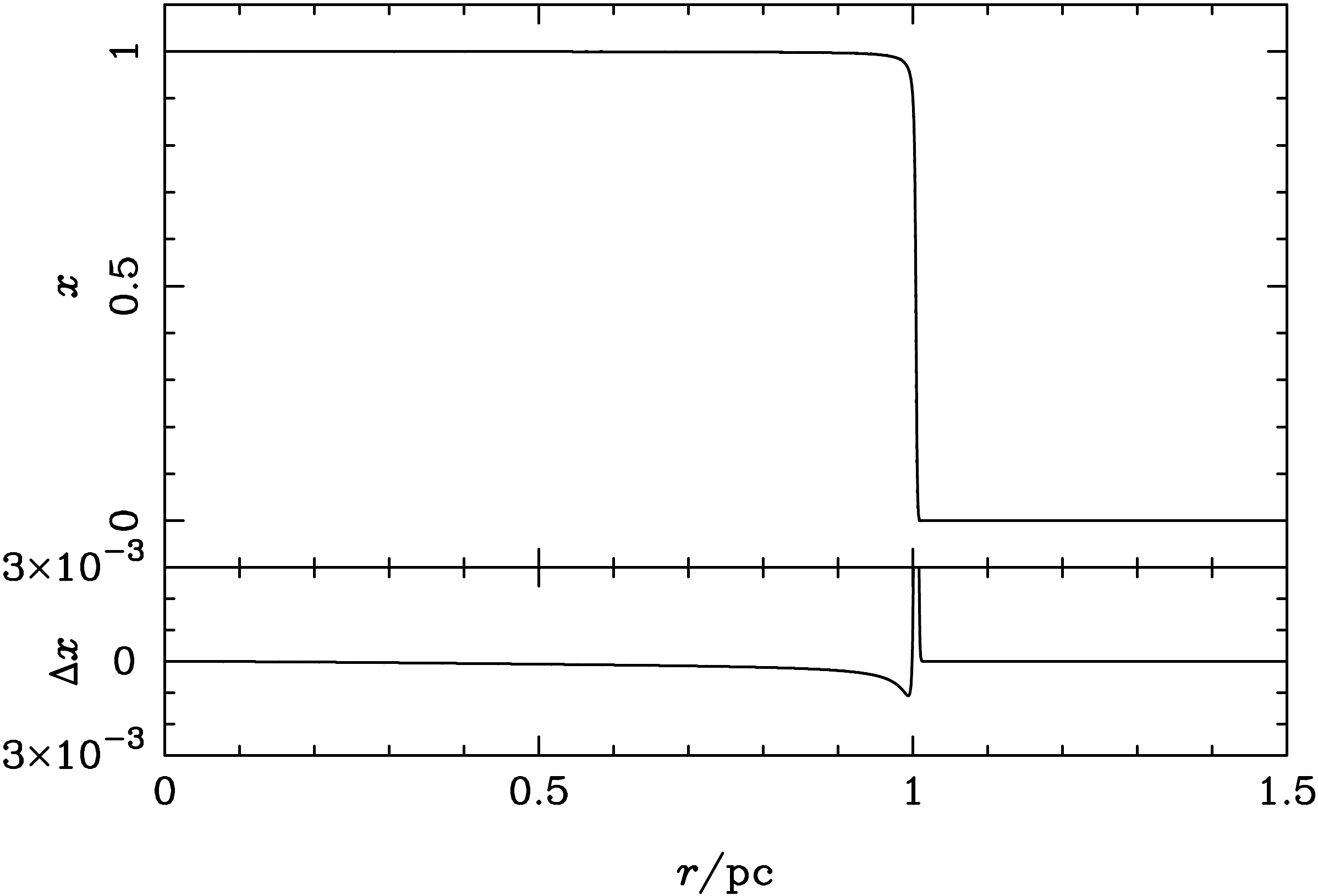} &
\includegraphics[width=6cm,angle=270]{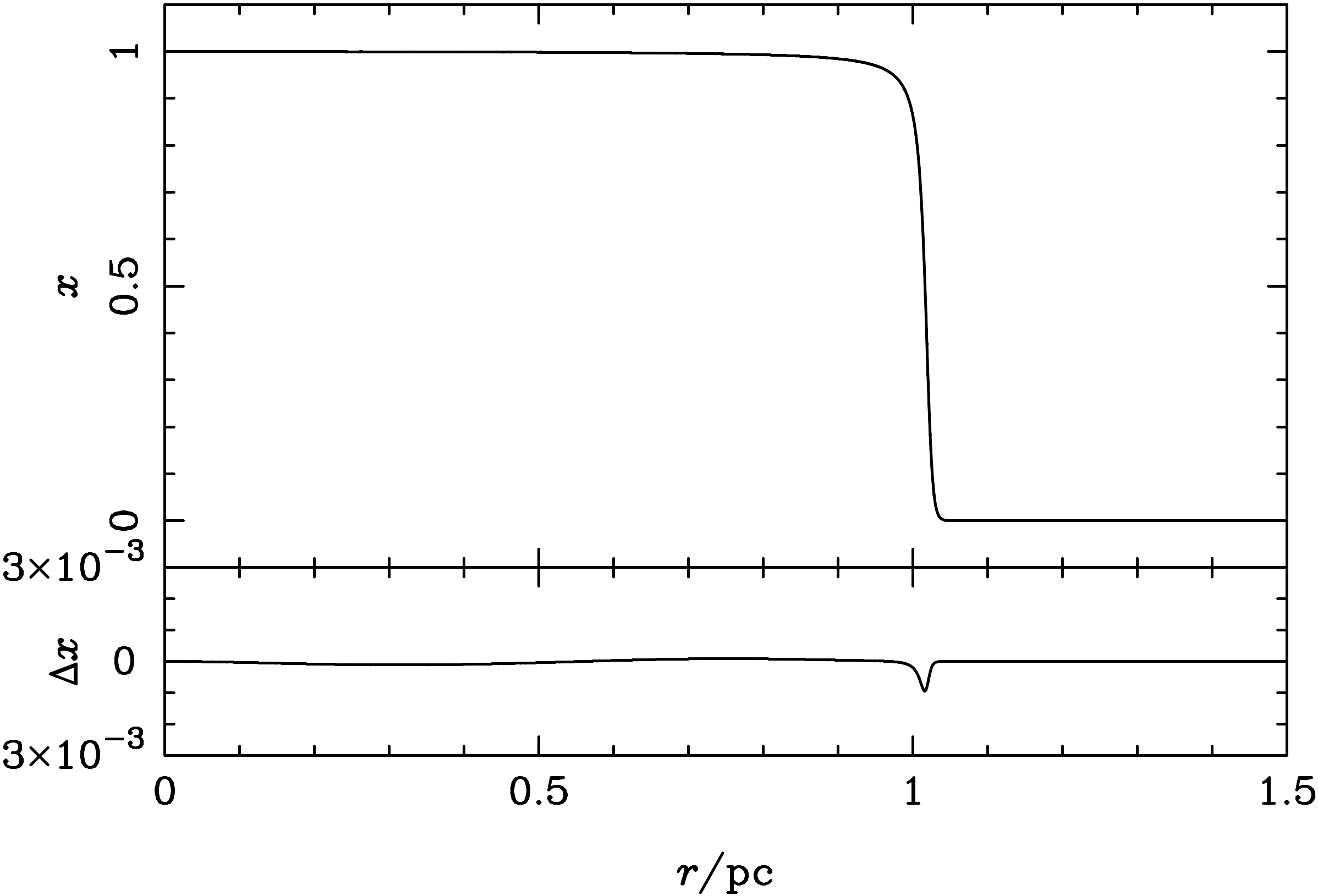} \\
\includegraphics[width=6cm,angle=270]{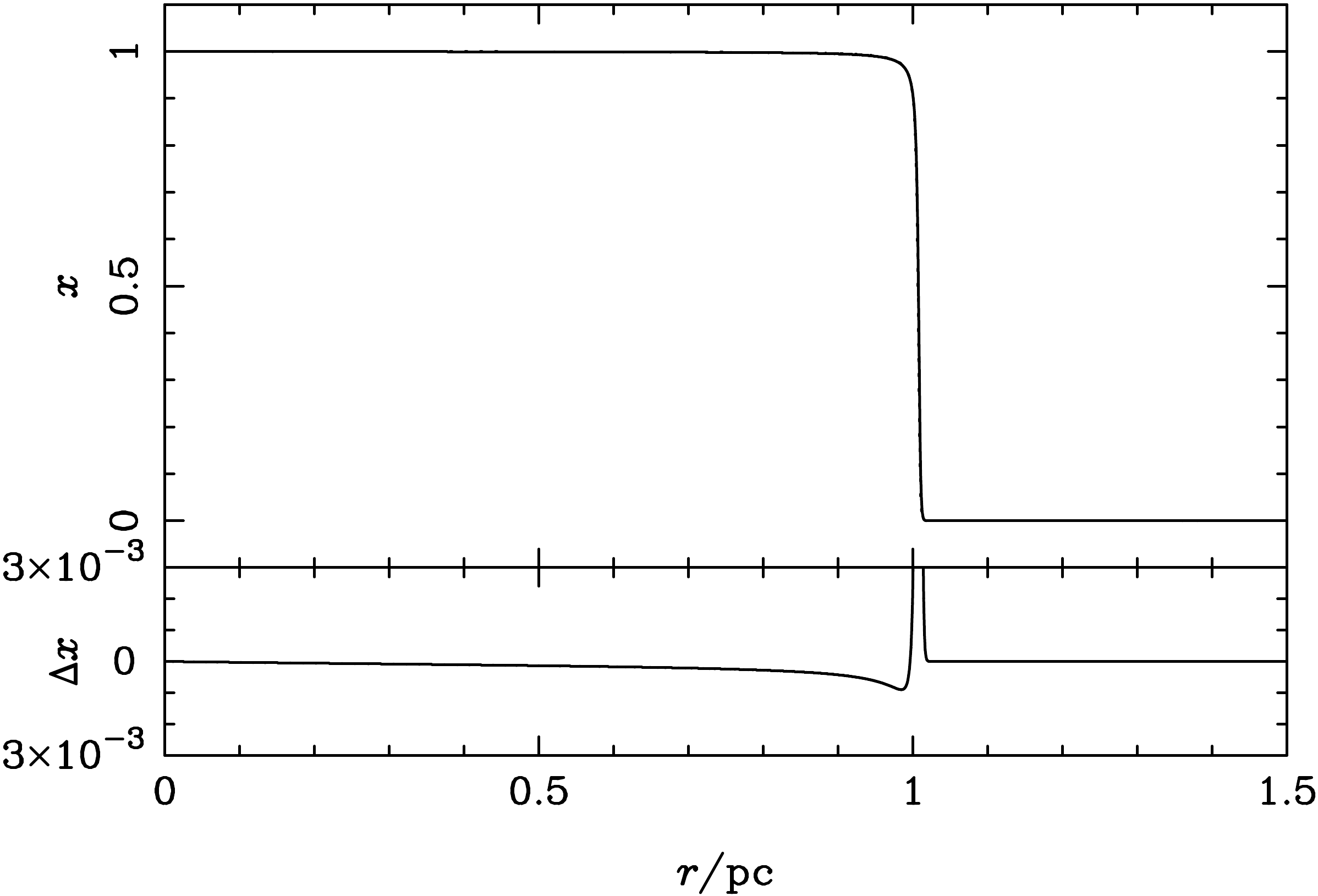} &
\includegraphics[width=6cm,angle=270]{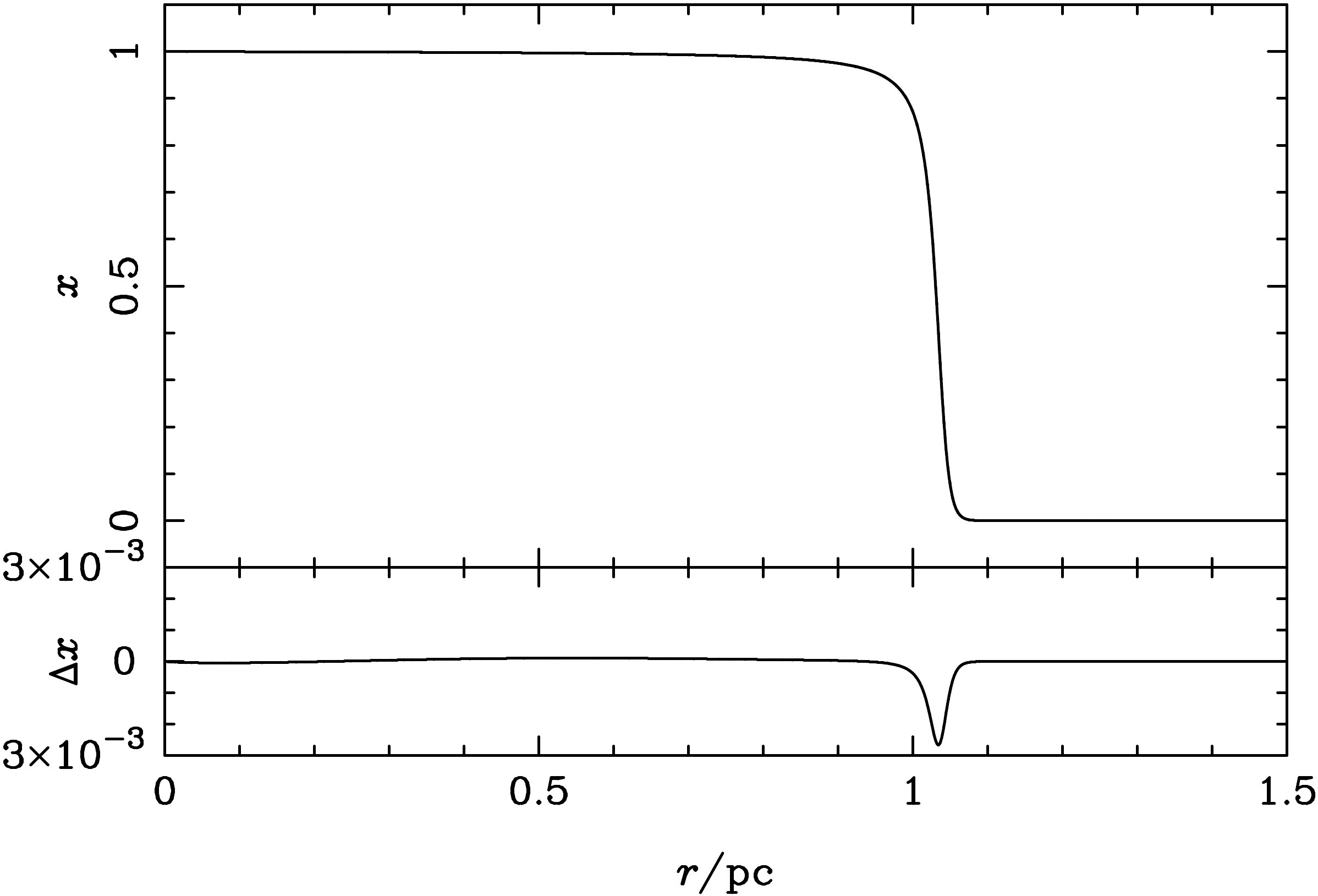} \\
\includegraphics[width=6cm,angle=270]{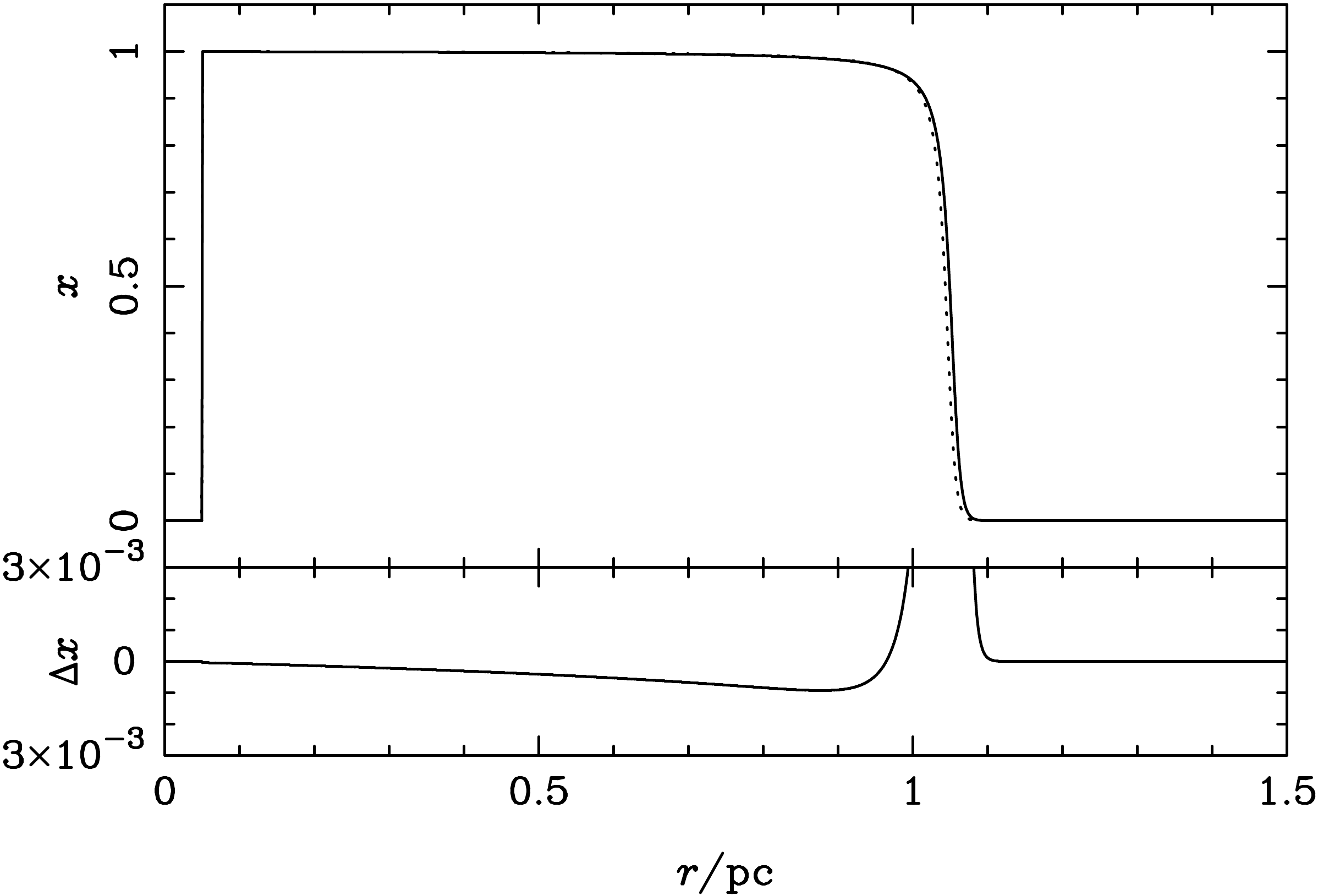} &
\includegraphics[width=6cm,angle=270]{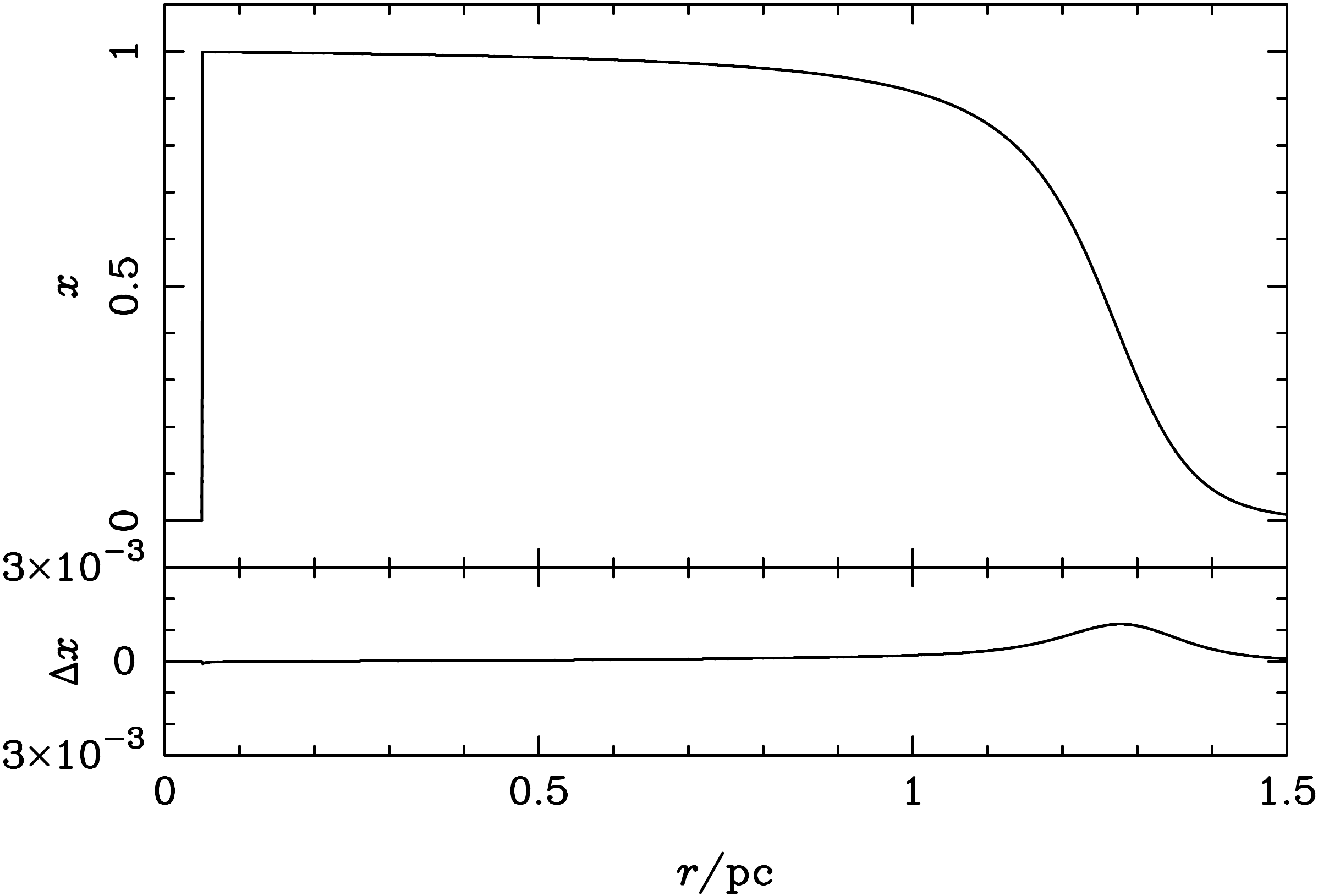} \\
\end{tabular}
\caption{H{\sc\,ii} region structures.  Left column, with equal
  absorption coefficients, right column with diffuse coefficient 6
  times direct.  Top row uniform density, second row $r^{-1}$ outside
  a small core, last row $r^{-2}$ outside 0.05 core.  Solid line is
  ionization fraction for full transfer solution, dotted is ionization
  fraction for OTS solution -- in almost all cases, the ionization
  fractions are essentially identical.  The lower panel in the plots
  shows $x_{\rm full} - x_{\rm OTS}$, with the scale set to show
  differences internal to the H{\sc\,ii} region.}
\label{f:ionize}
\end{figure*}

\begin{figure*}
\begin{tabular}{cc}
\includegraphics[width=6cm,angle=270]{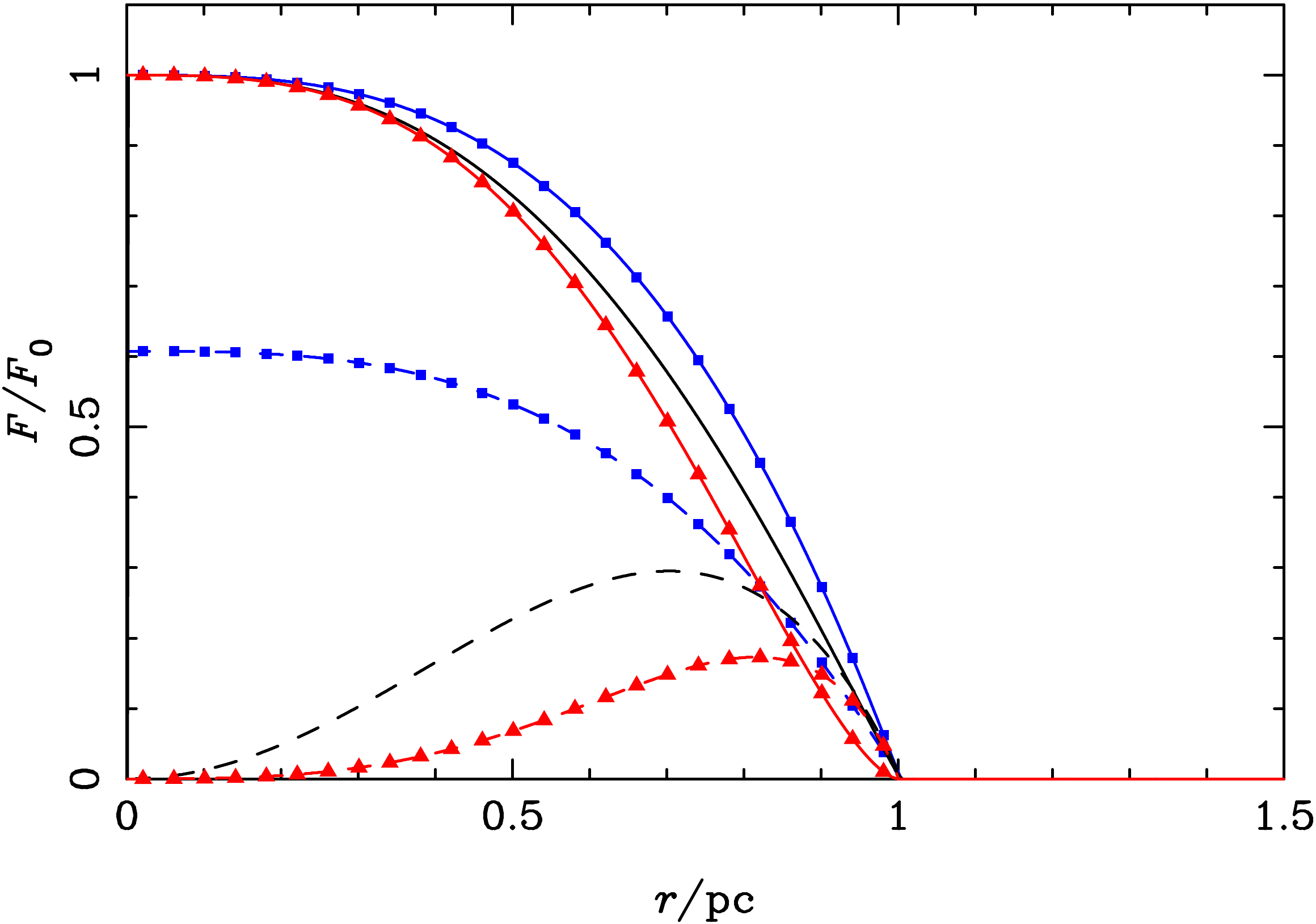} &
\includegraphics[width=6cm,angle=270]{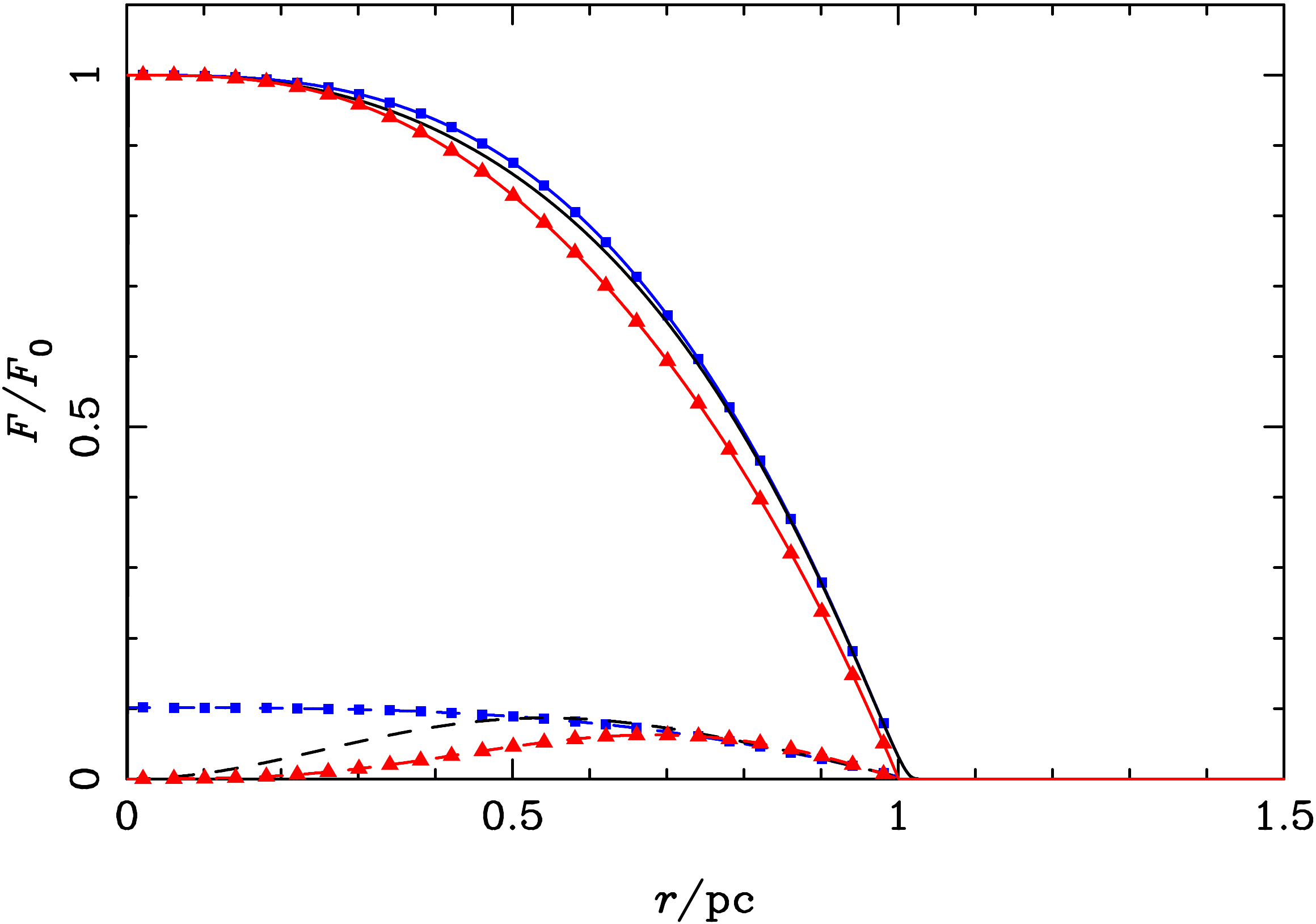} \\
\includegraphics[width=6cm,angle=270]{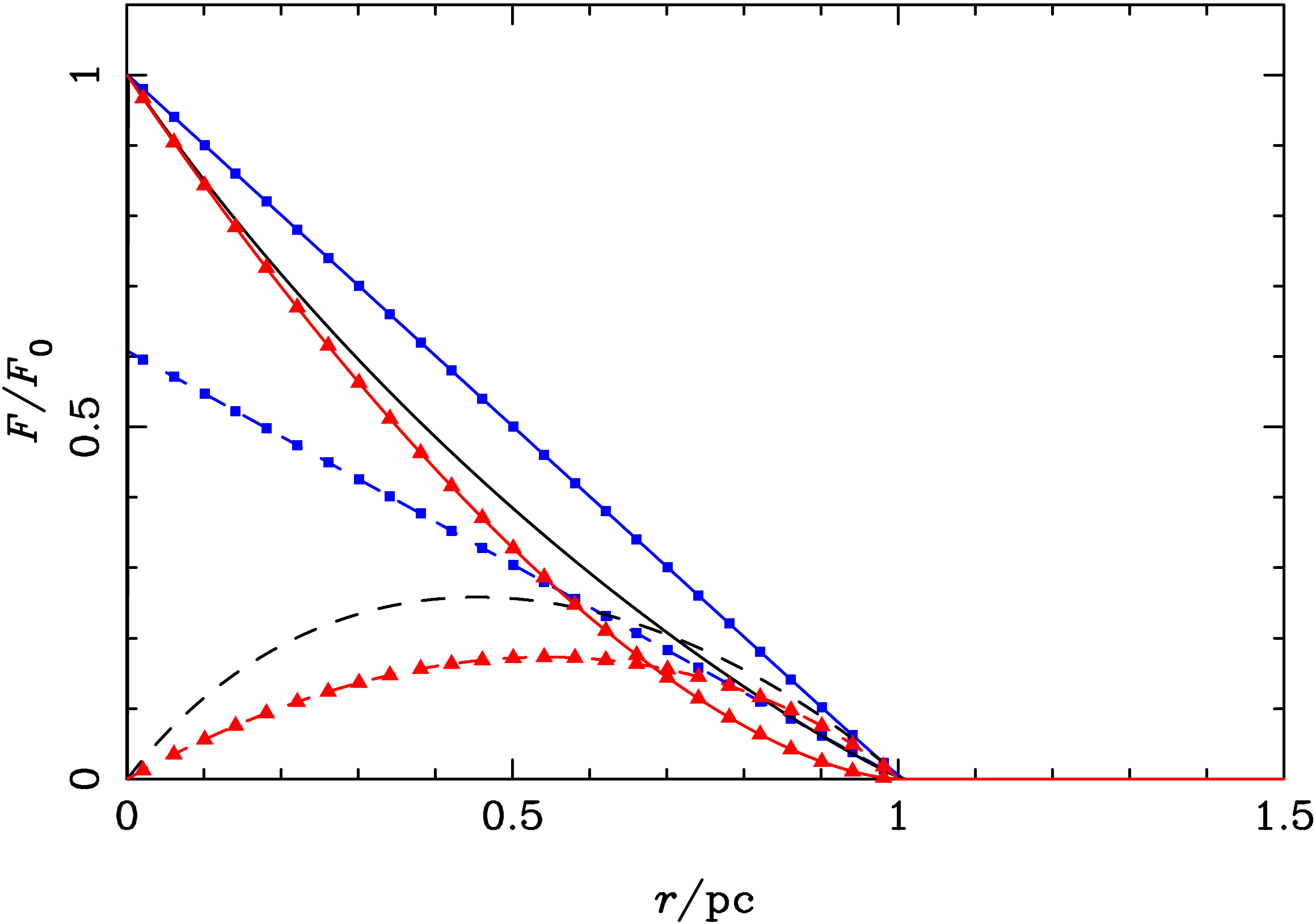} &
\includegraphics[width=6cm,angle=270]{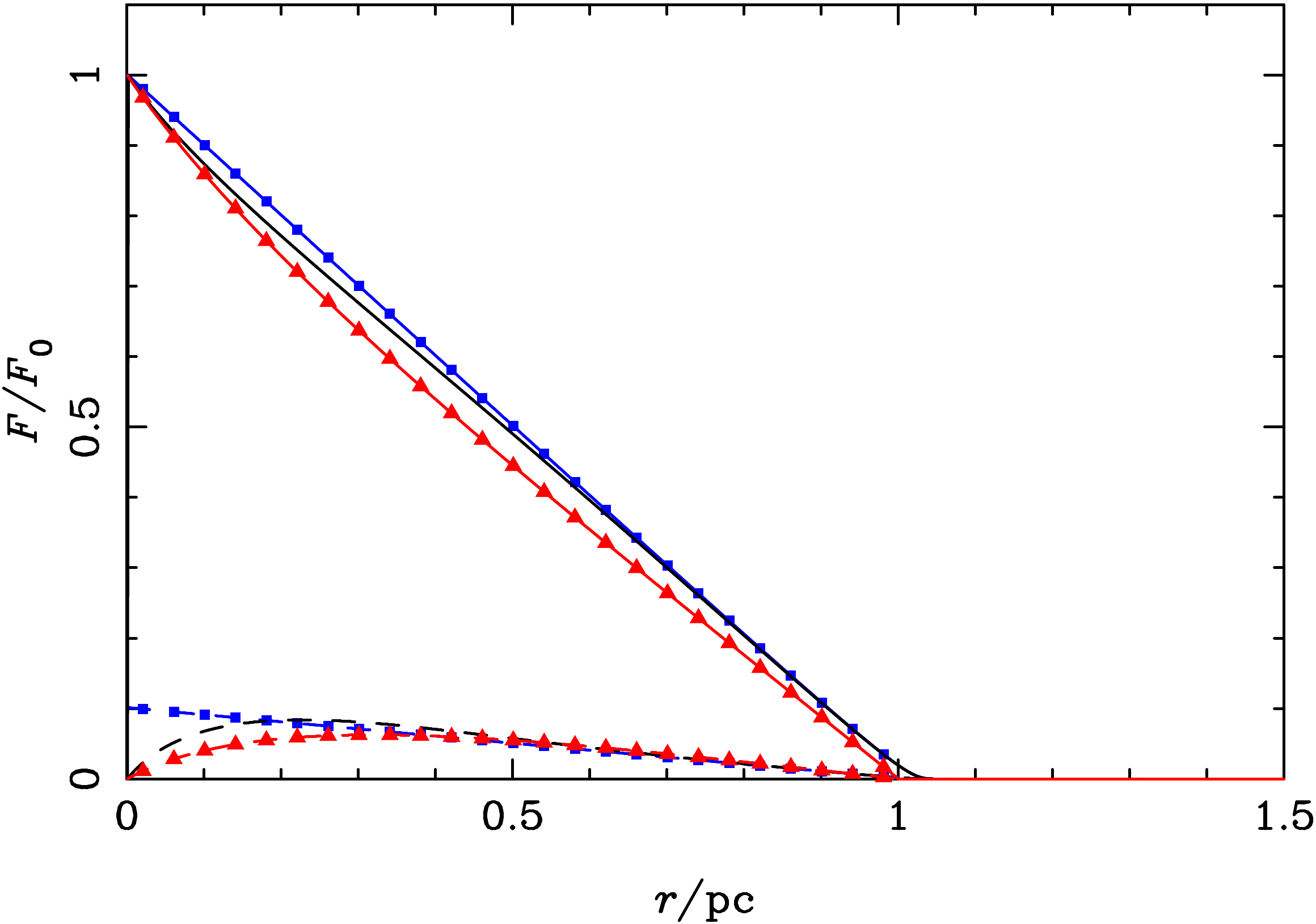} \\
\includegraphics[width=6cm,angle=270]{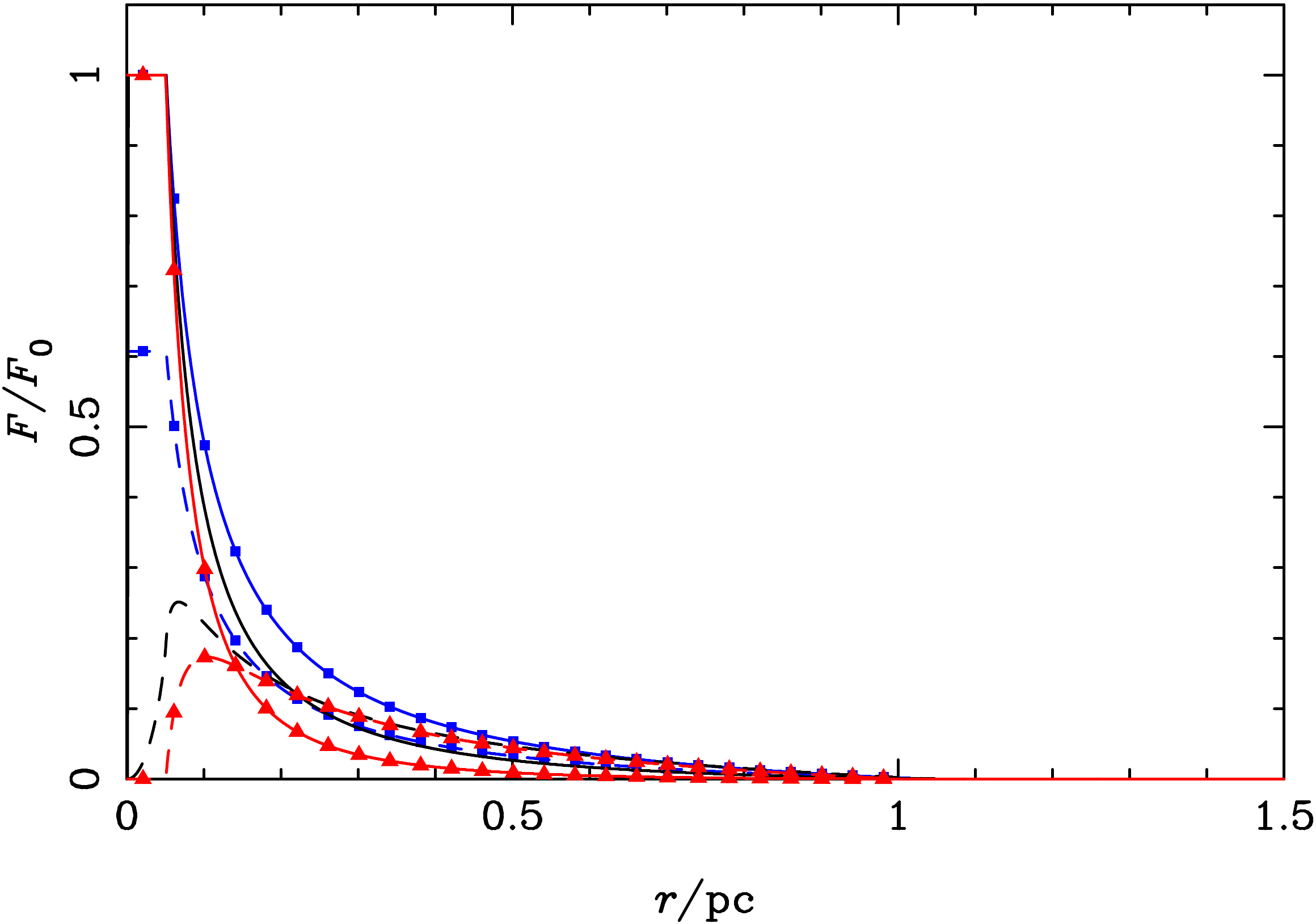} &
\includegraphics[width=6cm,angle=270]{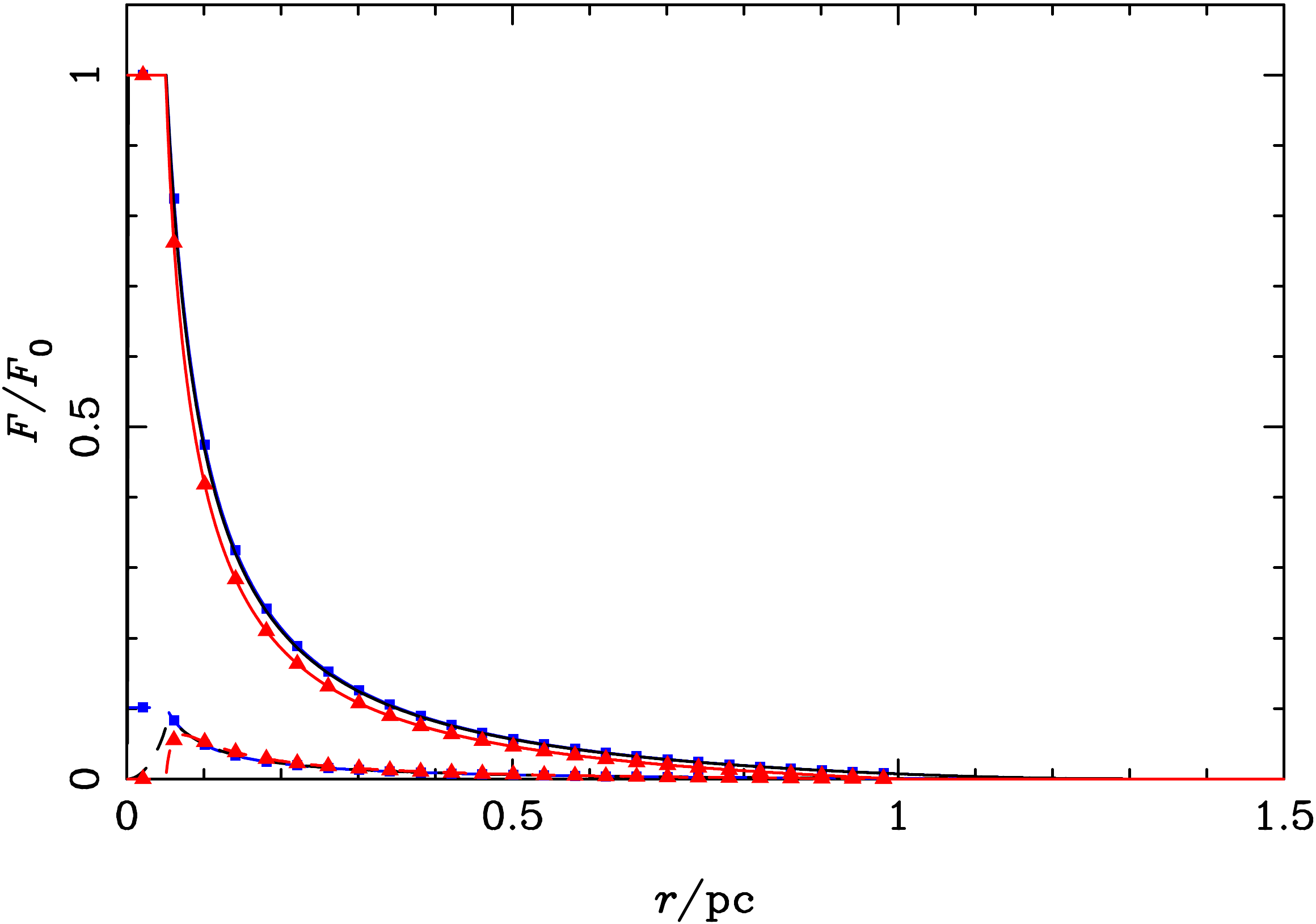} \\
\end{tabular}
\caption{H{\sc\,ii} region structures.  Left column, with equal
  absorption coefficients, right column with diffuse coefficient 6
  times direct.  Top row uniform density, second row $r^{-1}$ outside
  a small core, last row $r^{-2}$ outside 0.05 core.  Solid is direct
  photon rate, dashed is $4\pi r^2$ times the local diffuse intensity.
  OTS results have square markers, the approximation of
  \protect\cite{ritze05} has triangular markers; the detailed transfer
  solution is plain.  In general, the exact results lie between the
  two approximations; in lower right plot, the OTS and exact results
  are closely coincident.}
\label{f:radiat}
\end{figure*}

\begin{figure*}
\begin{tabular}{cc}
\includegraphics[width=6cm,angle=270]{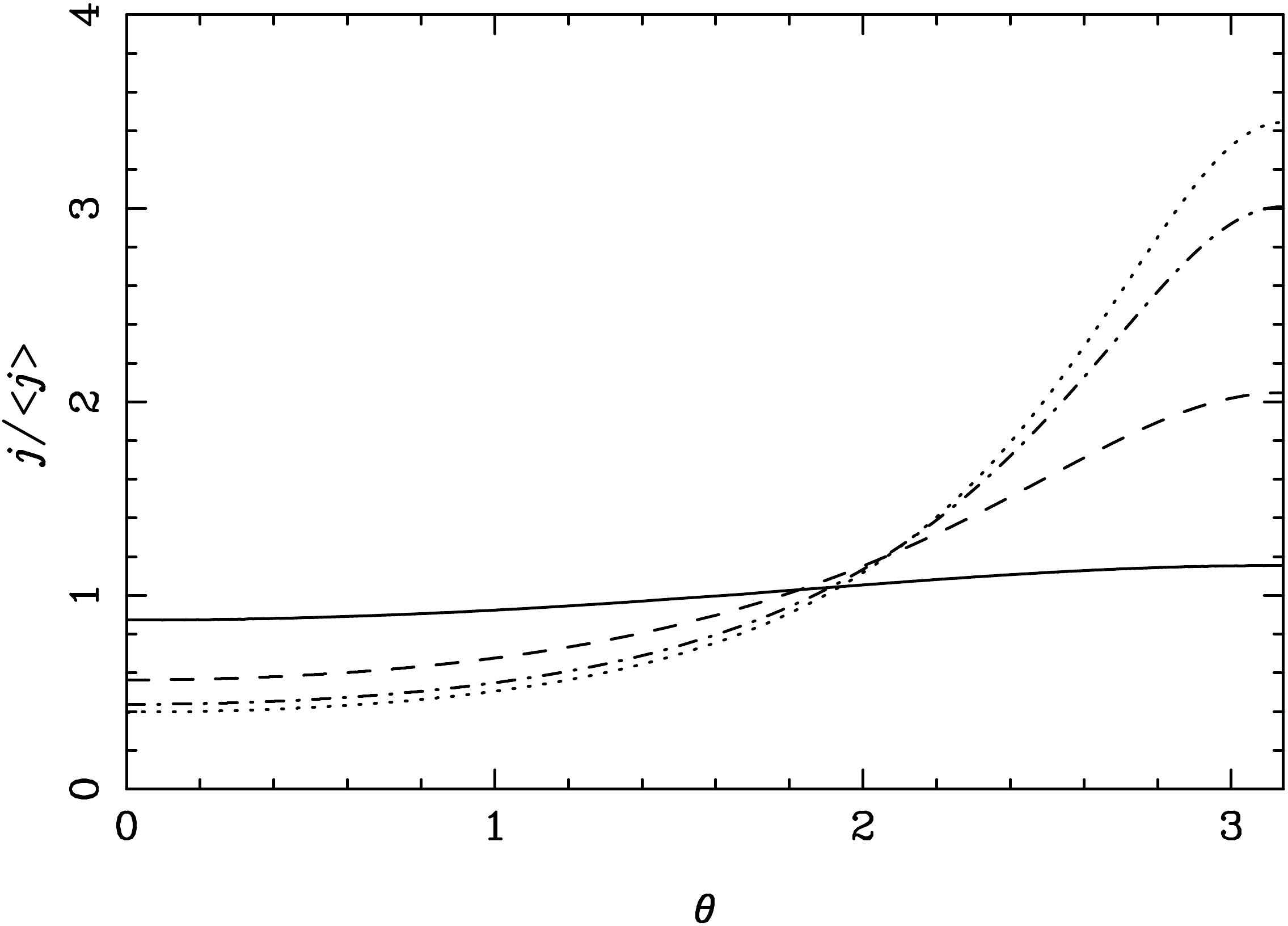} &
\includegraphics[width=6cm,angle=270]{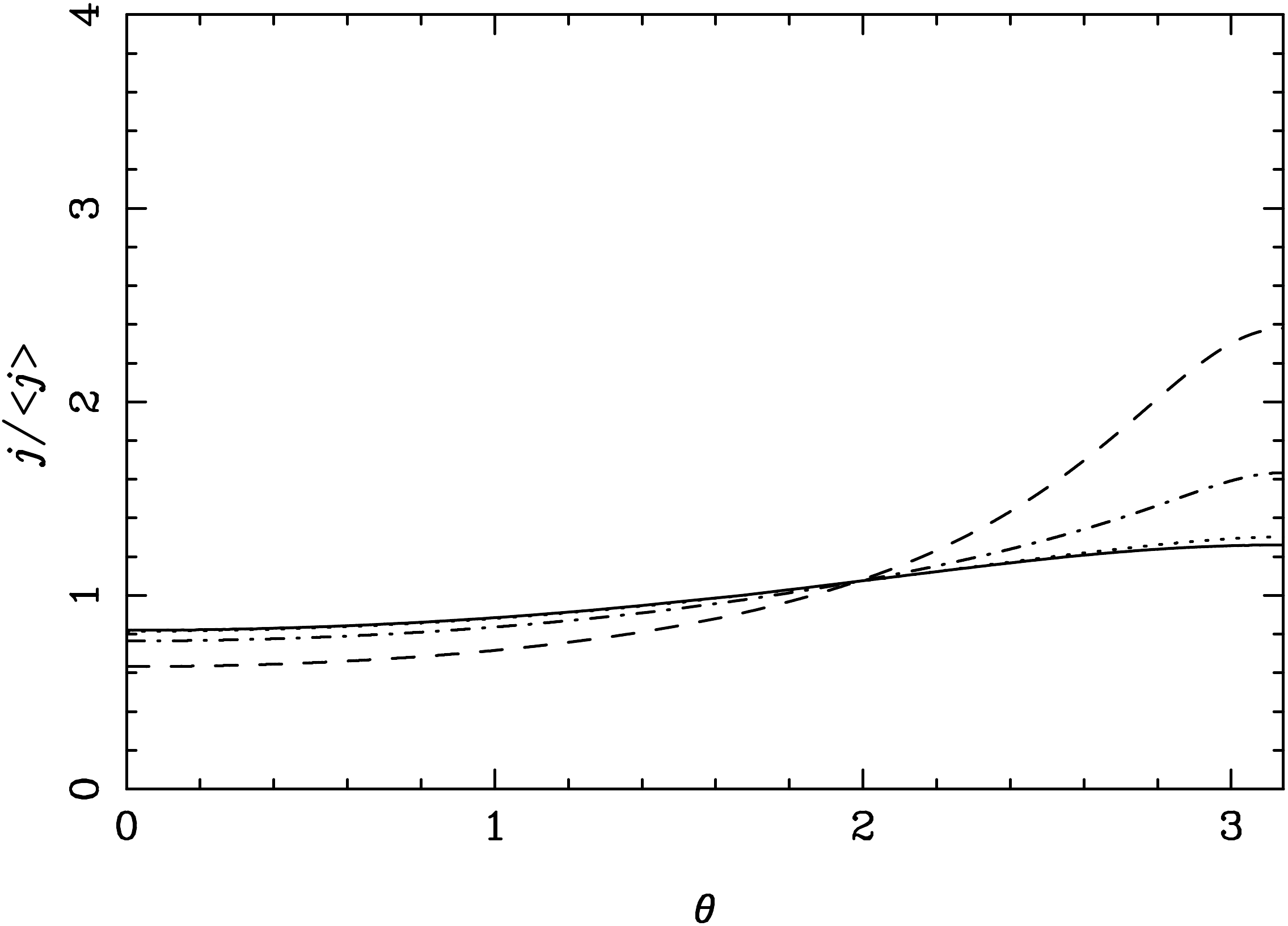} \\
\includegraphics[width=6cm,angle=270]{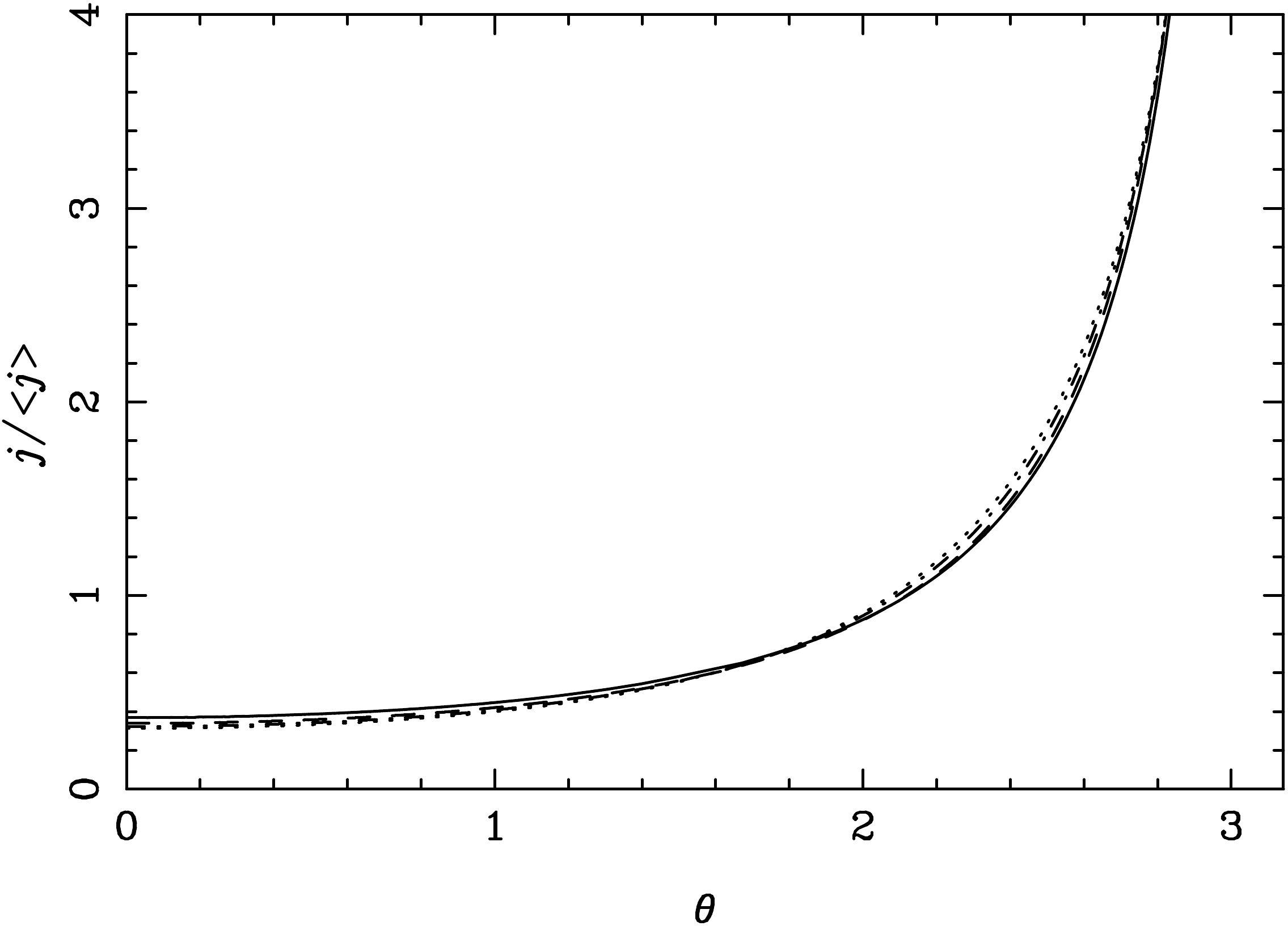} &
\includegraphics[width=6cm,angle=270]{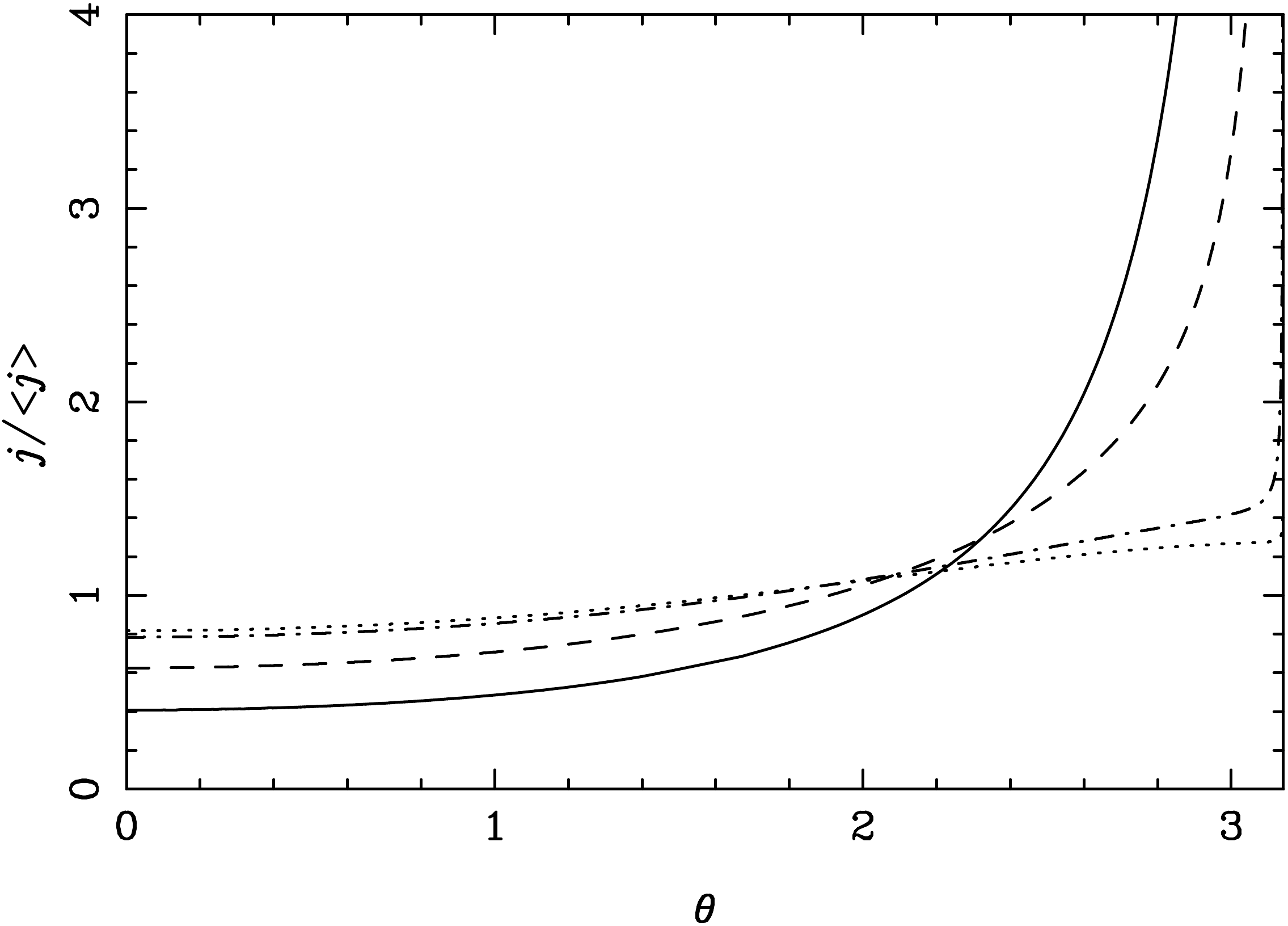} \\
\includegraphics[width=6cm,angle=270]{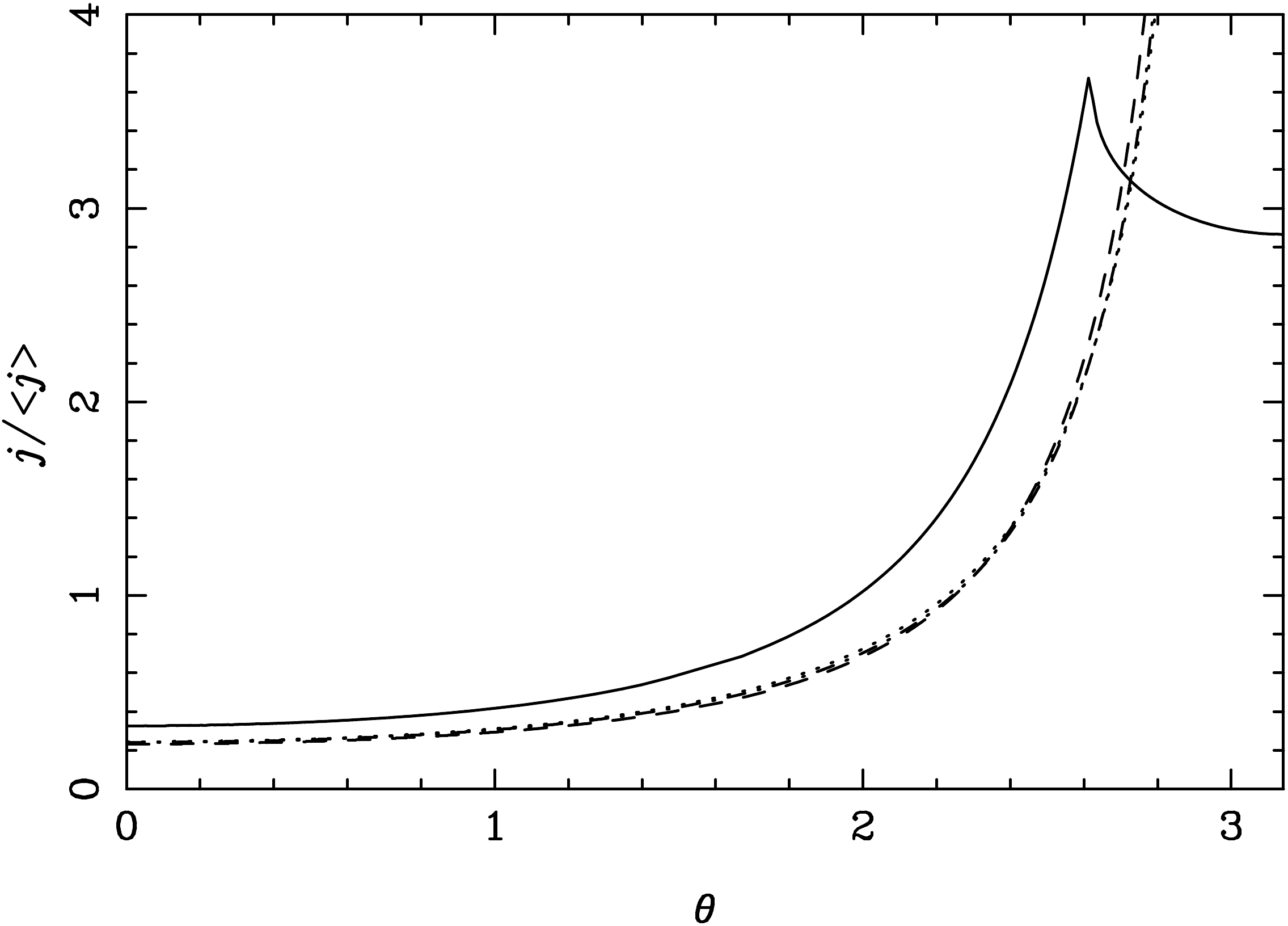} &
\includegraphics[width=6cm,angle=270]{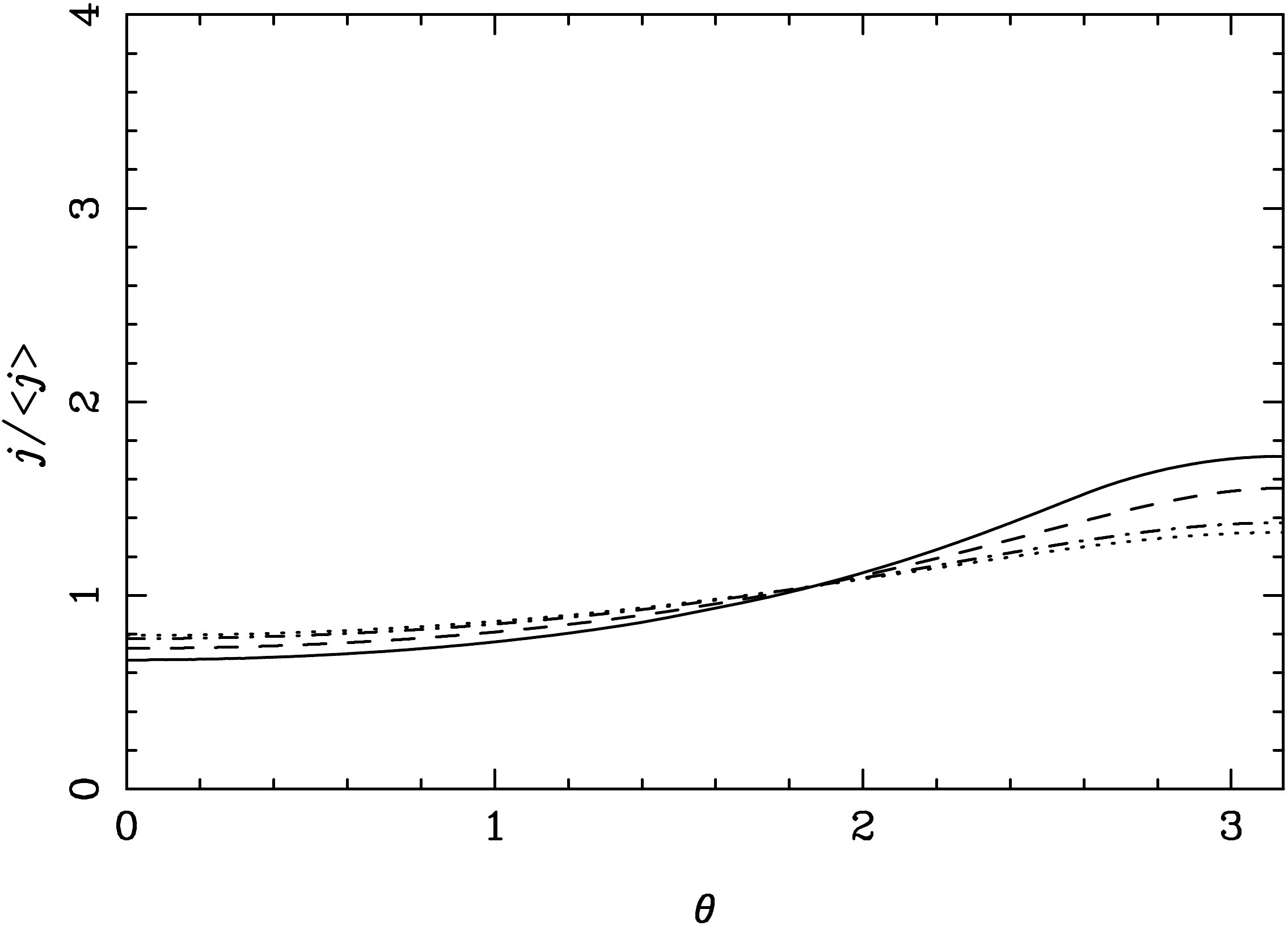} \\
\end{tabular}
\caption{Polar diagrams for the radiation at $0.1$ (solid), $0.5$
  (dashed), $0.8$ (dot-dashed) and $0.9$ (dotted) $\times r_{\rm S}$.
  Left column, with equal absorption coefficients, right column with
  diffuse absorption coefficient 6 times that for direct photons.  Top
  row uniform density, second row $r^{-1}$ outside a small core, last
  row $r^{-2}$ outside 0.05 core.}
\label{f:polar}
\end{figure*}

\begin{figure*}
\begin{tabular}{cc}
\includegraphics[width=6cm,angle=270]{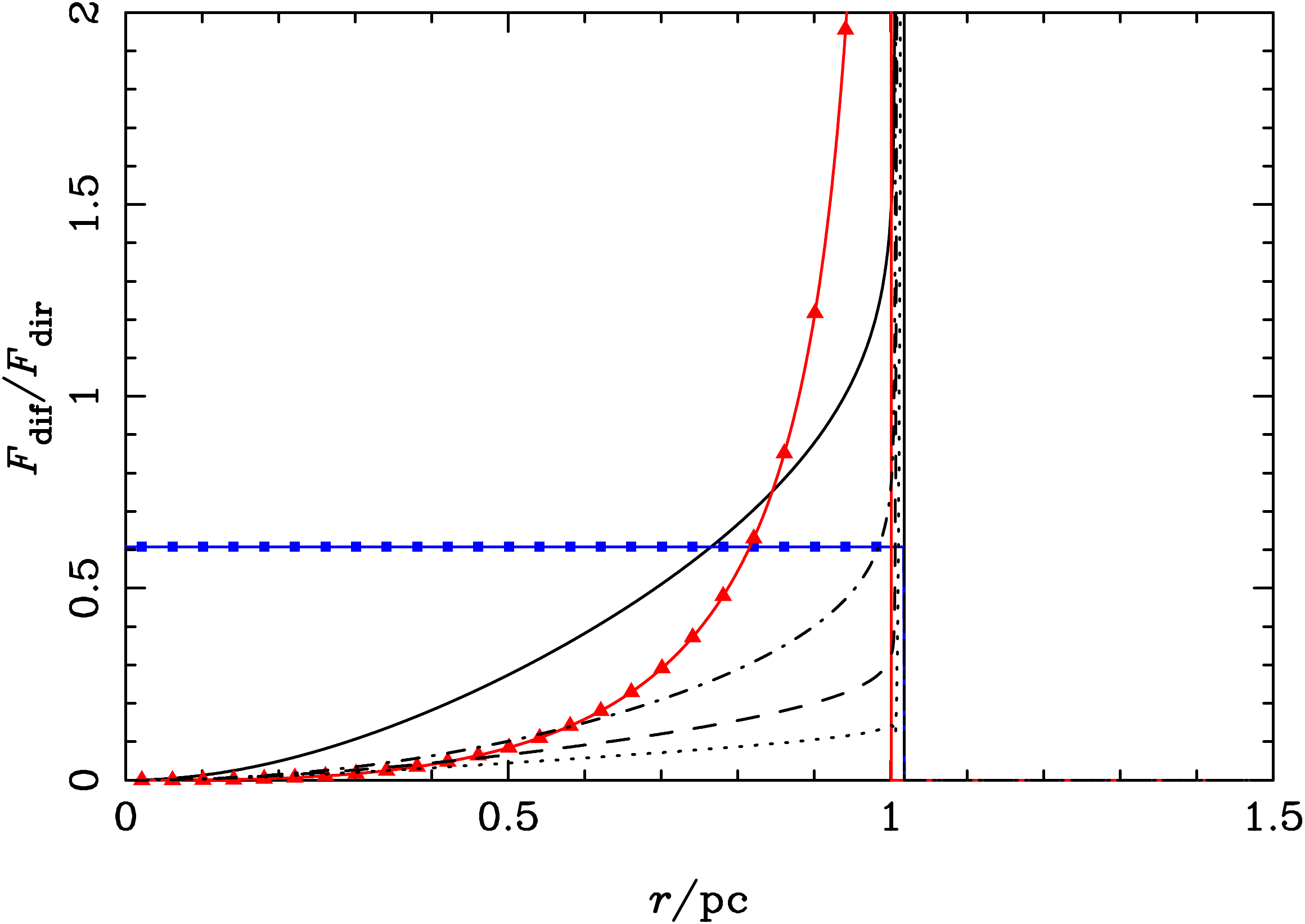} &
\includegraphics[width=6cm,angle=270]{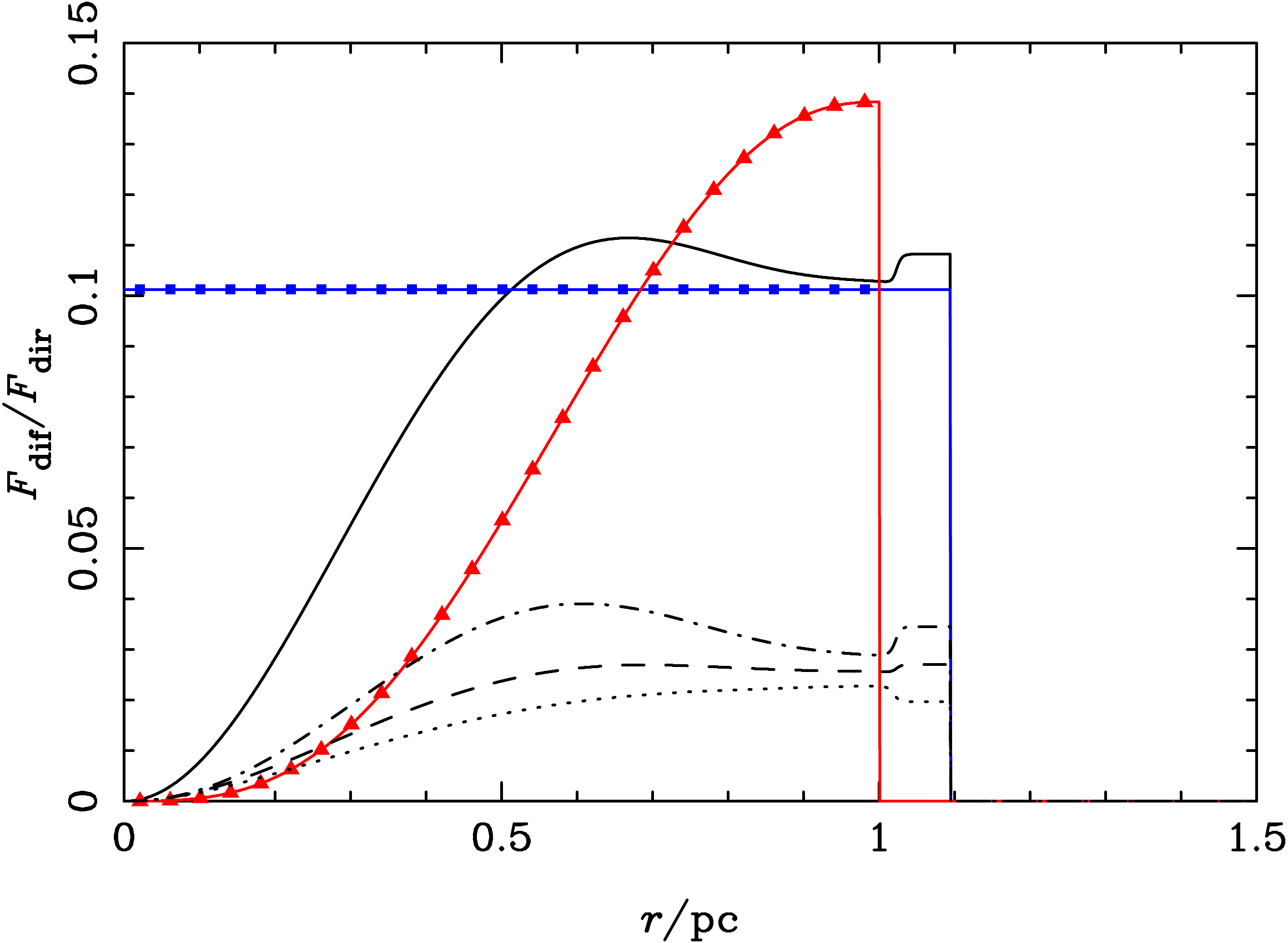} \\
\includegraphics[width=6cm,angle=270]{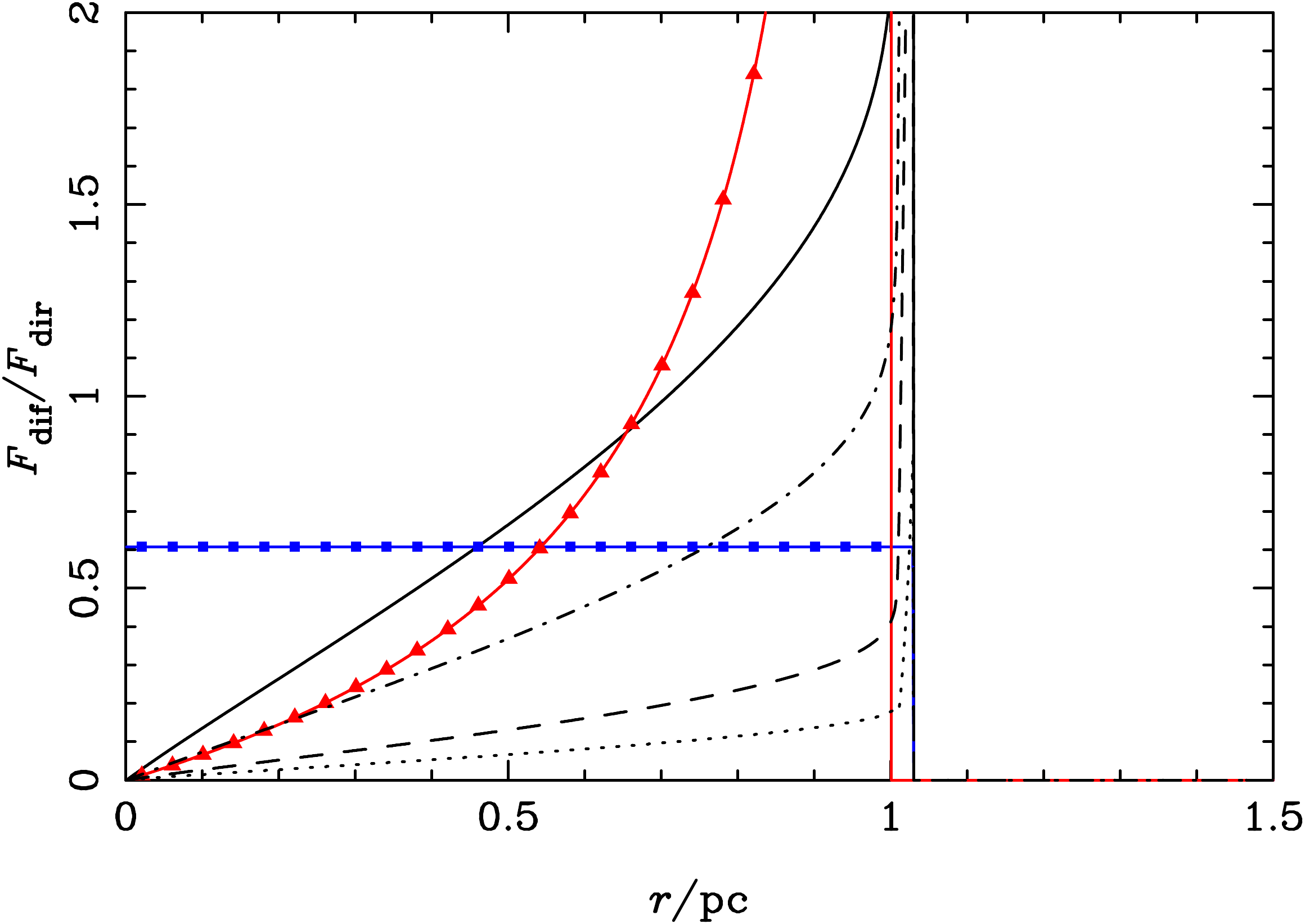} &
\includegraphics[width=6cm,angle=270]{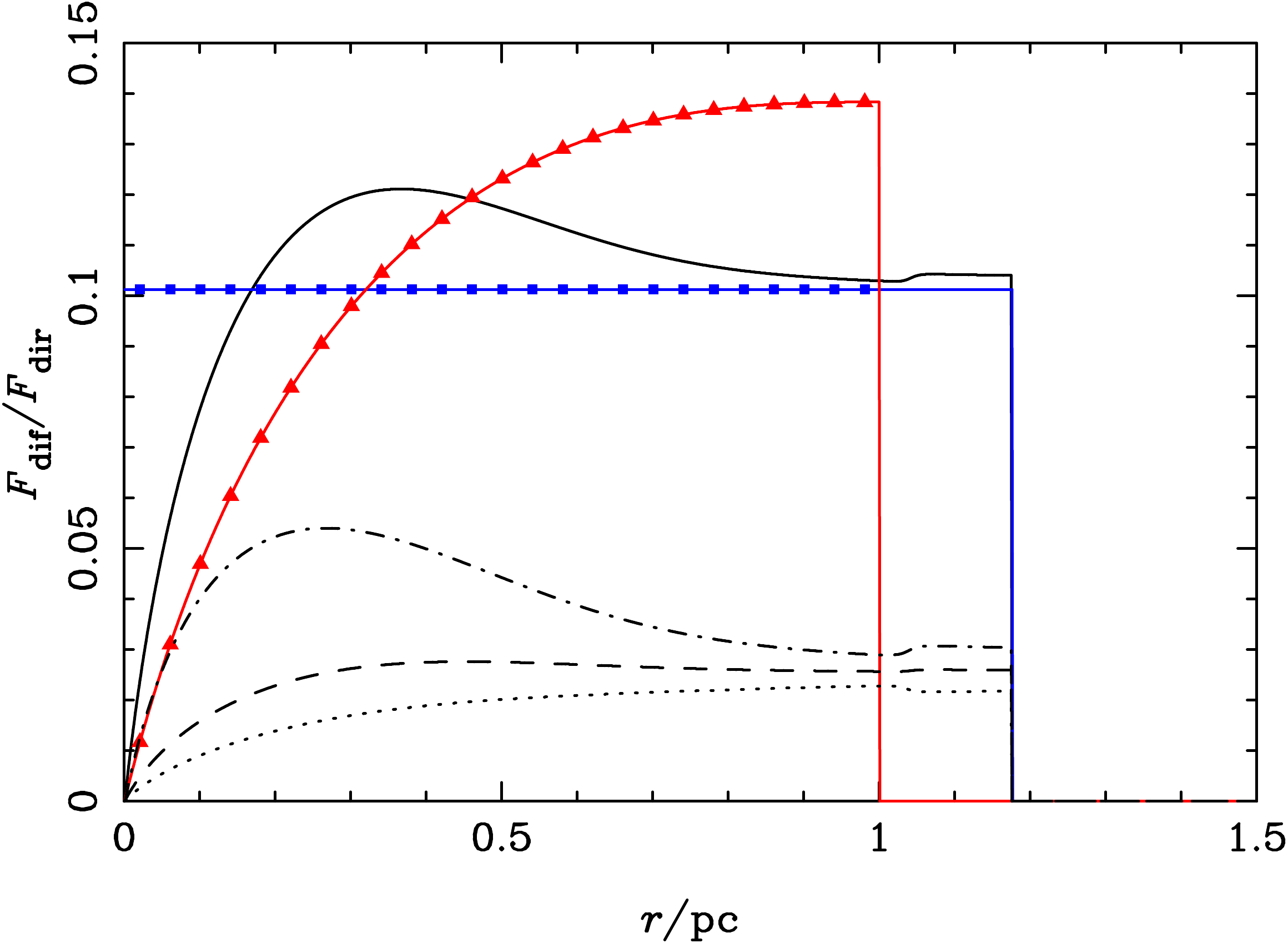} \\
\includegraphics[width=6cm,angle=270]{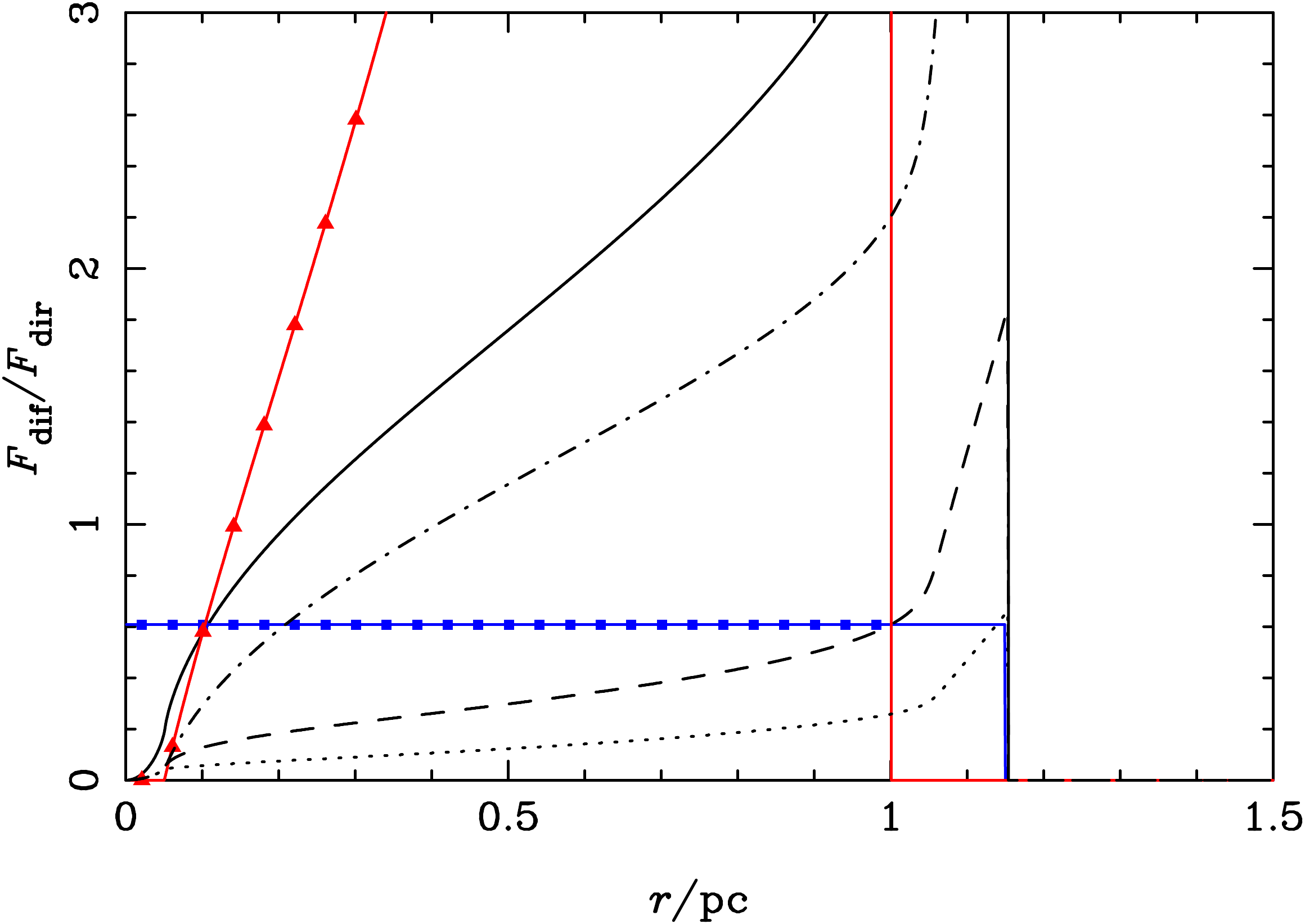} &
\includegraphics[width=6cm,angle=270]{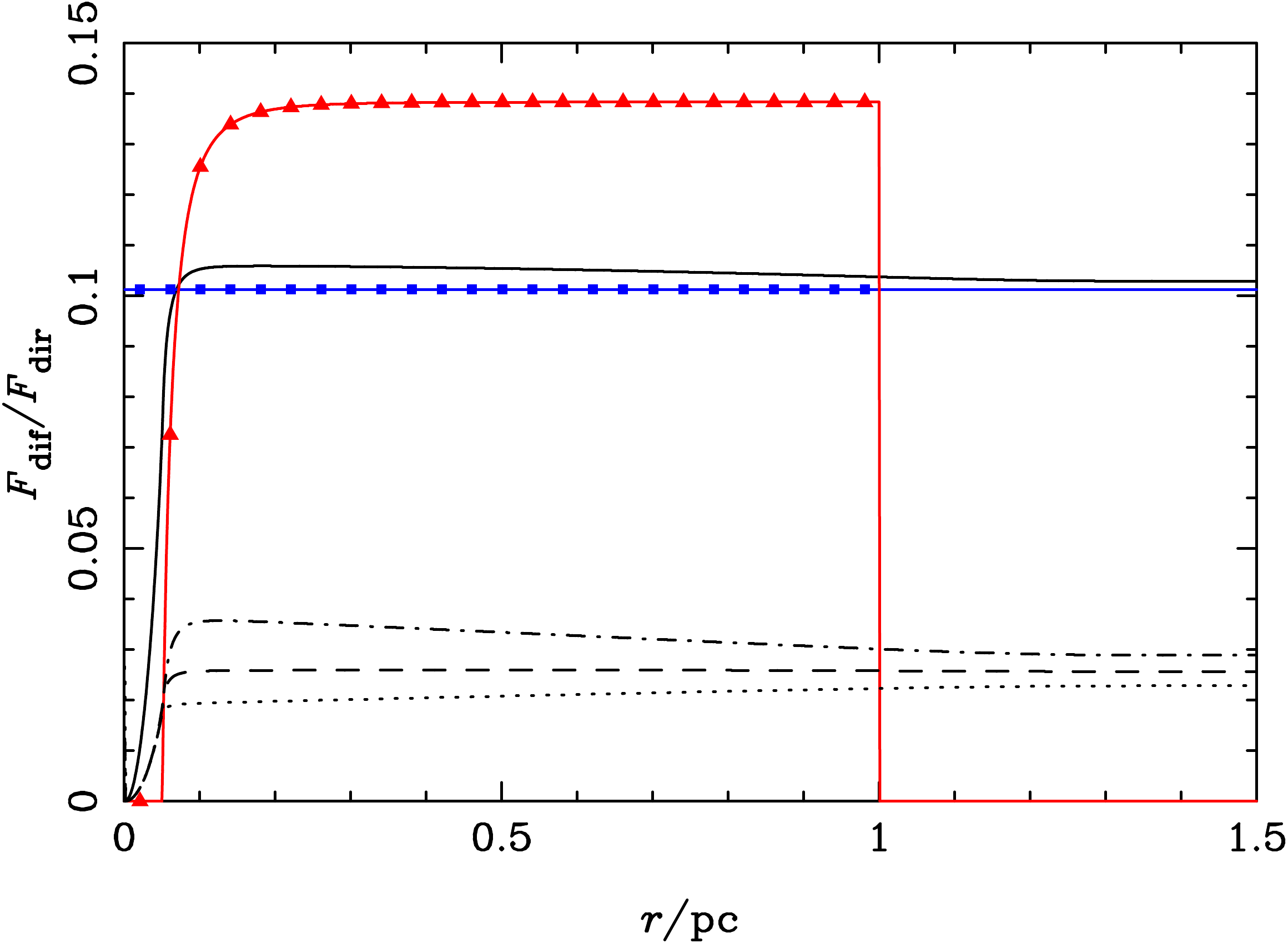} \\
\end{tabular}
\caption{Diffuse field components as a fraction of the direct field
  for the full-transfer solution.  Left column, with equal absorption
  coefficients, right column with diffuse coefficient 6 times direct.
  Top row uniform density, second row $r^{-1}$ outside a small core,
  last row $r^{-2}$ outside 0.05 core.  Solid is the total diffuse
  energy density, dashed is the flux incident tangentially, dot-dashed
  is the outward flux, dotted is the inward flux.  Square and
  triangular markers on solid lines show the total diffuse energy
  density estimated by the OTS and Ritzerveld approximations,
  respectively.}
\label{f:radang}
\end{figure*}

\begin{figure*}
\begin{tabular}{cc}
\includegraphics[width=6cm,angle=270]{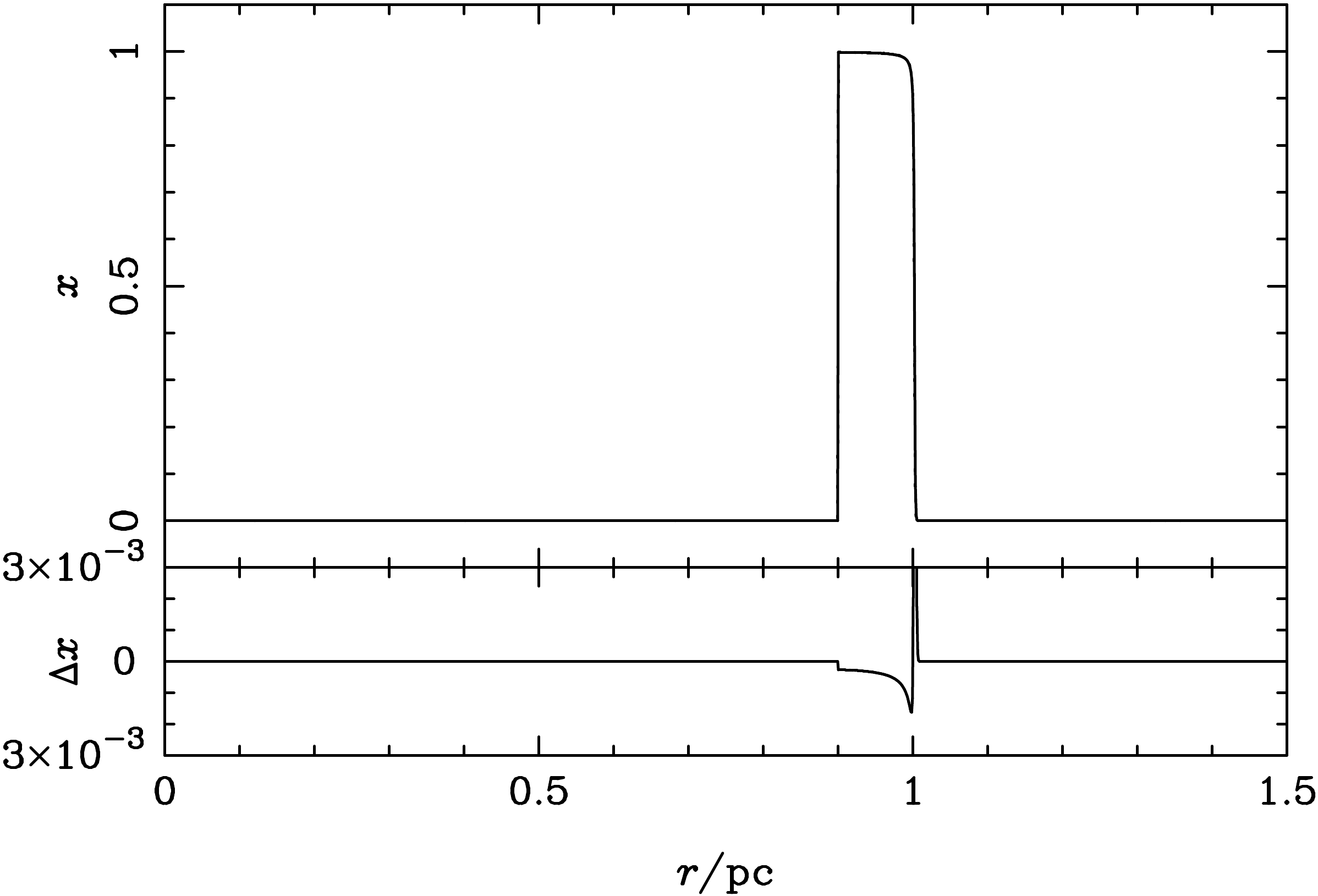} &
\includegraphics[width=6cm,angle=270]{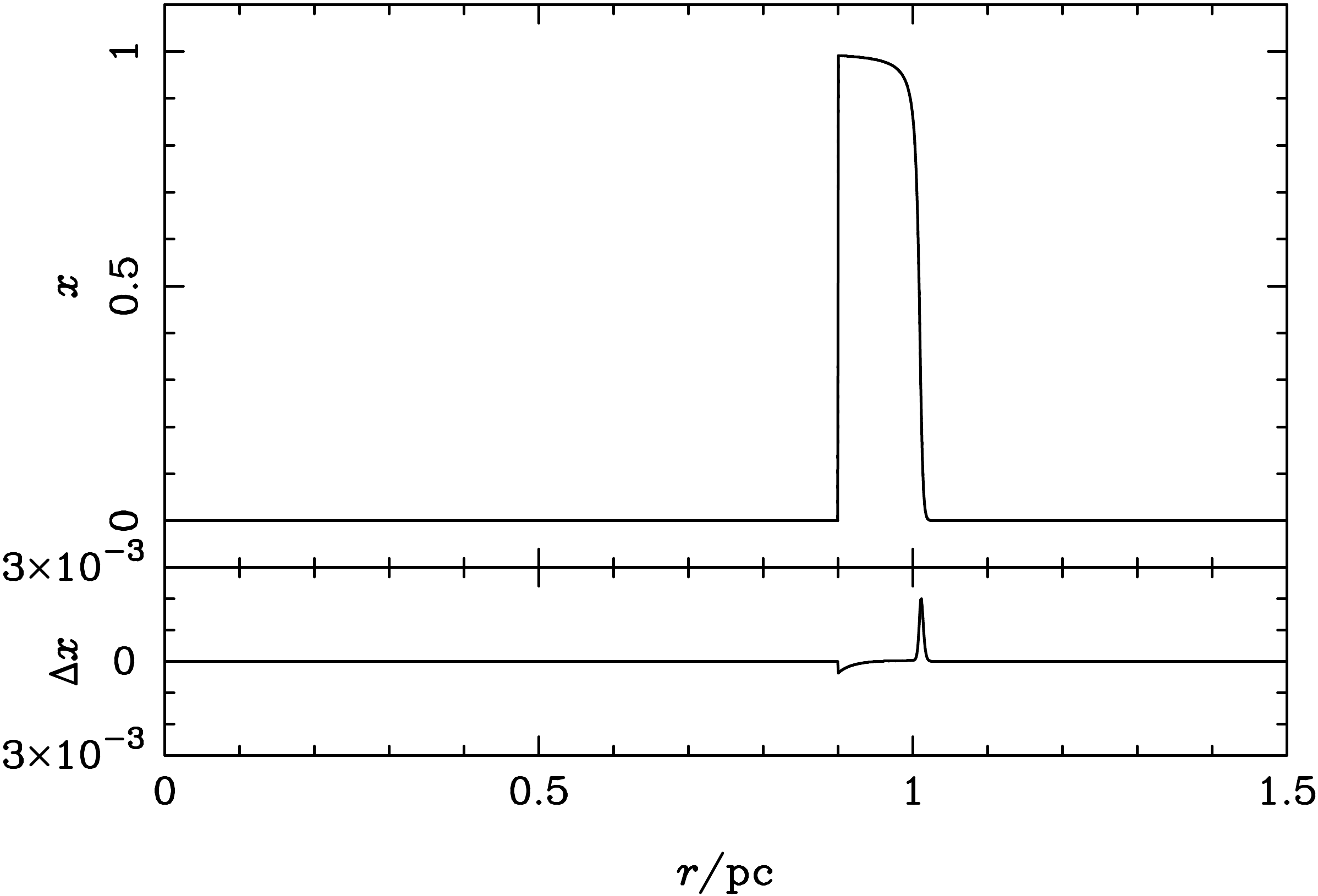} \\
\includegraphics[width=6cm,angle=270]{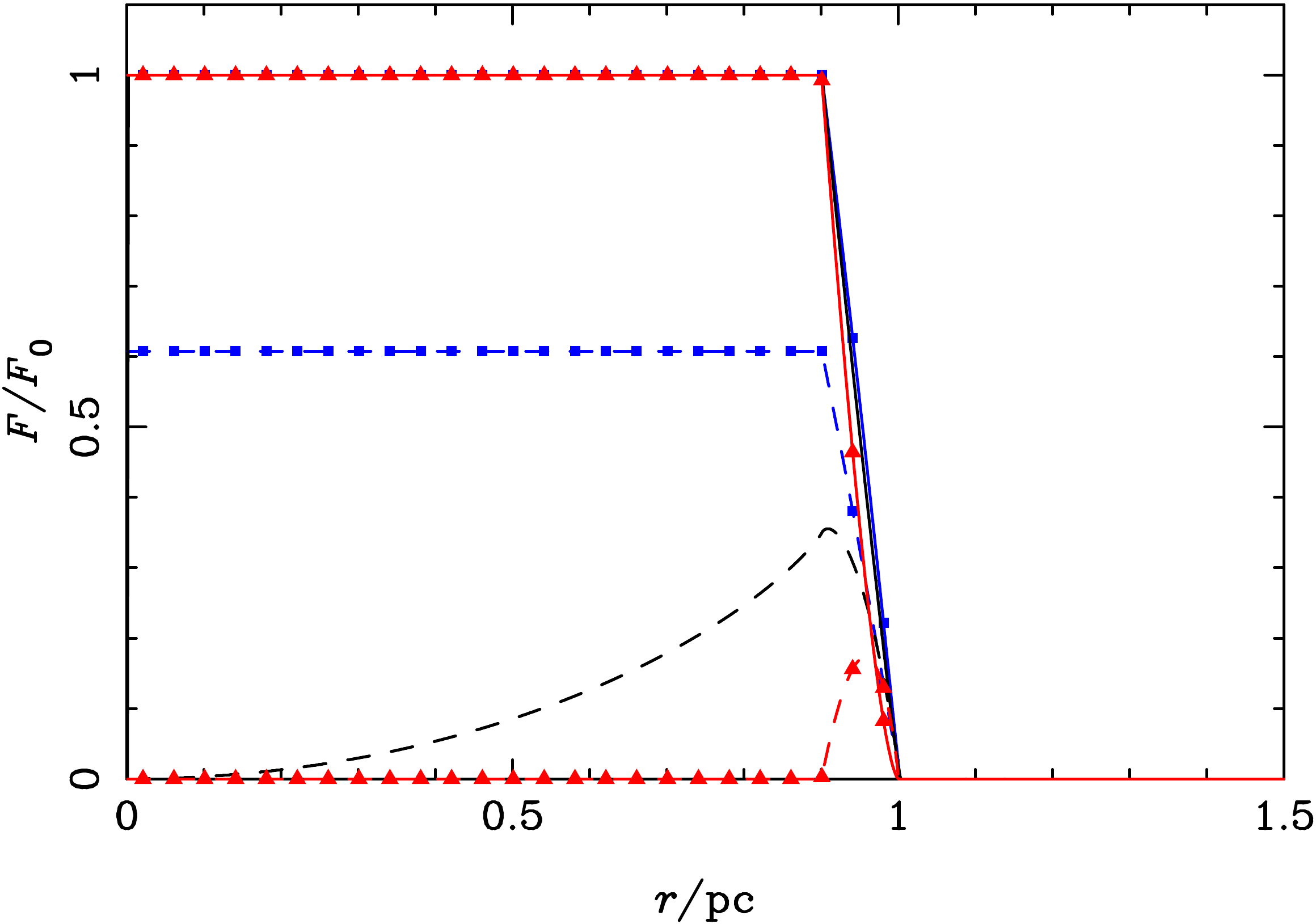} &
\includegraphics[width=6cm,angle=270]{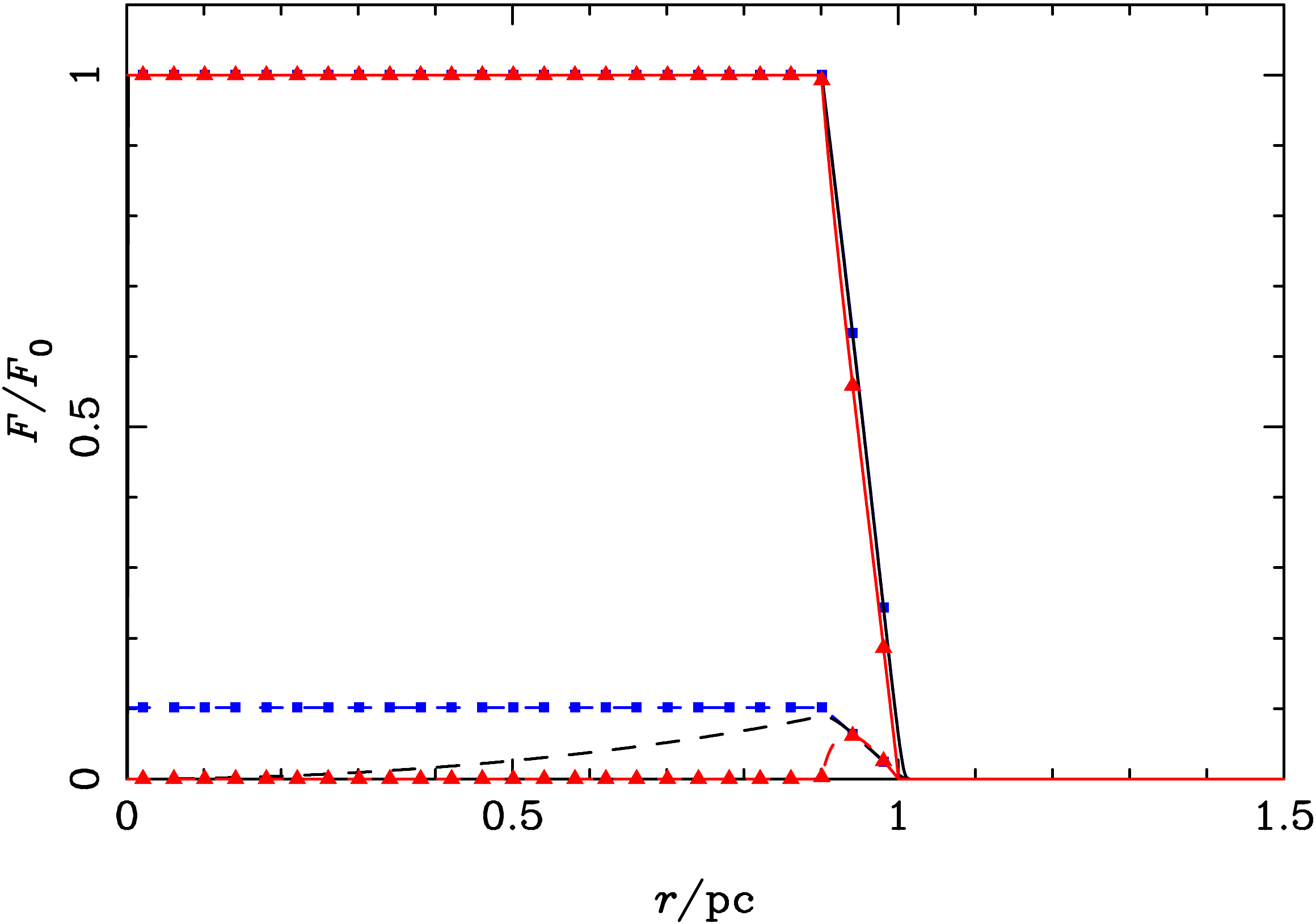} \\
\includegraphics[width=6cm,angle=270]{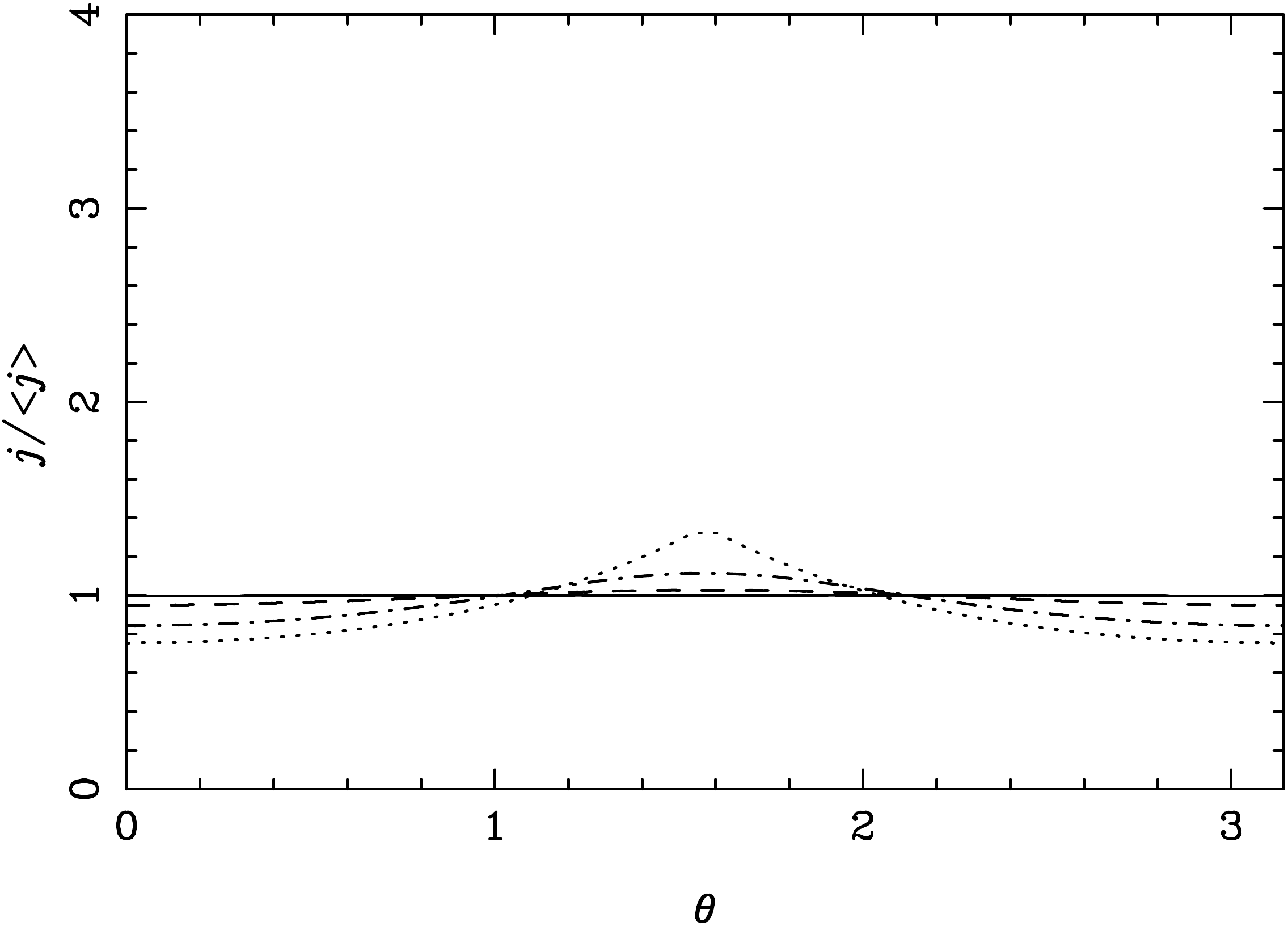} &
\includegraphics[width=6cm,angle=270]{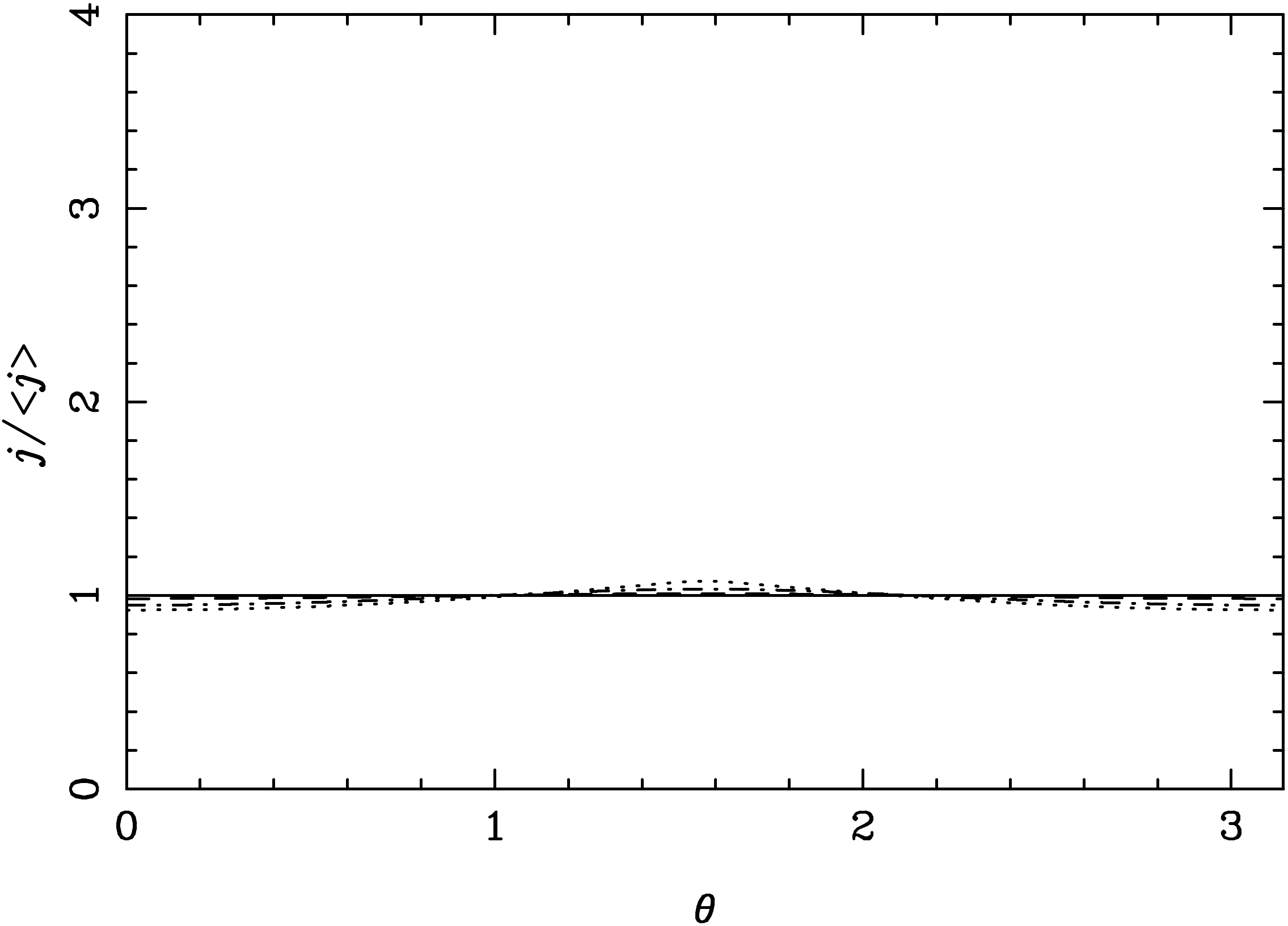} 
\end{tabular}
\caption{H{\sc\,ii} region structure and polar diagram for a thin
shell with zero density out to 90 per cent of the Str\"omgren radius
and uniform density beyond.  Left column, with equal absorption
coefficients, right column with diffuse absorption coefficient 6 times
direct.  The line style and colours of the curves have the same
meanings as defined in previous figures.}
\label{f:shell}
\end{figure*}

In this section, we present the results of the diffuse radiation
solutions using a variety of numerical schemes, compared to the full
transfer solutions.  The full transfer solutions have been discretized
on a grid with spatial resolution of $\Delta = 0.001r_{\rm S}$ out to
$2r_{\rm S}$.  

In Figure~\ref{f:ionize}, we show the ionization fraction structures
for the OTS and exact transfer solutions.  These are seen to be very
close to each other in all cases, excpet for close to the ionization
front, particularly in the case of equal absorption coefficients.  In
the case of an $r^{-2}$ density distribution, the outer edge of the
H{\sc\,ii} region is significantly broader.  This is true for both the
OTS and the full transfer solutions.  The structure of the outer
region of the ionized bubble in the $r^{-2}$ case is highly sensitive
to details of the internal structure, as can be seen from the relation
\begin{equation}
L_\star = 4\pi\alpha_{\rm B}n_{\rm c}^2r_{\rm c}^4 (r_{\rm
c}^{-1}-r_{\rm S}^{-1}),
\end{equation}
so discretization errors at the 1 per cent level lead to changes in
the outer radius of the ionized bubble $r_{\rm S}/r_{\rm c}$ times
greater, i.e.\ 20 per cent for the parameters chosen by
\cite{ritze05}.  In reality, it is likely that the ionized region
would either be trapped in the core of the density distribution or
extend to infinity; if the static solution were to be found in a
region of such rapidly decreasing density, it is likely that the high
sensitivity in its position would result in dynamical instability.

In Figure~\ref{f:radiat} we show the direct and diffuse radiation
intensities.  As expected, the diffuse field intensity is higher in
the case of equal absorption coefficients.  Overall, the complete
integration gives results roughly intermediate between the OTS and
Ritzerveld approximations.  

The OTS approximation overestimates the diffuse field intensity in the
core of the nebula, while the Ritzerveld approximation underestimates
this.  The direct field in the outer part of the nebula is reduced in
the outer parts of the nebula in the Ritzerveld approximation,
primarily as a result of the lower ionization in the core.  For the
case of $r^{-2}$, the assumption that $x=1$ within the Str\"omgren
sphere means that this approximation also does not capture the
broadening of the ionization front.

In Figure~\ref{f:polar}, we show the angular distribution of the
diffuse radiation at various positions within the nebula.  Where the
absorption coefficients of direct and diffuse radiation are the same,
the diffuse radiation field is also beamed strongly forward through
much of the nebula, because it is principally emitted in the dense gas
in the core.  Hence, although the radiation energy may be distributed
with the form characteristic of recombination emission, effects of a
directed radiation field such as shadowing should still be in
evidence.  For the uniform density case with diffuse absorption
coefficient six time larger than direct, the radiation field is most
strongly beamed at intermediate radii within the nebula: the
directionality is small at $r=0.1{\rm\,pc}$ for both uniform density
cases as this is close to the centre of symmetry, where as for the
higher absorption case the diffuse field is also symmetric near to the
edge of the nebula because the $\tau_{\rm dif} \sim 1$ surface samples
a relatively small degree of anisotropy in the direct field.

For the innermost radius of the $r^{-2}$ distribution with the higher
absorption coefficient for the direct field, the inner edge of the
photoionized gas can be clearly seen in the diffuse radiation field.

In Figure~\ref{f:radang}, we plot the diffuse radiation intensity as a
fraction of the direct radiation intensity as a function of radius for
our standard set of parameters.  These results complement the polar
diagrams at specific radii shown in Figure~\ref{f:polar}.  As we have
seen, in the case of equal absorption coefficients, the diffuse
intensity can be higher than the direct intensity.  However the ratio
of the {\em lateral}\/ diffuse flux to the direct flux, relevant to
the illumination of the tails of cometary globules, reaches at most 60
per cent at the edge of a $r^{-2}$ density distribution, and through
most of the nebula is closer to the 15 per cent figure derived by
\cite{canto98}.  In the case where the direct flux is less strongly
absorbed, all the diffuse flux components are close to 2.5 per cent of
the direct flux through almost all of the region; the outward-going
flux shows the most structure, but only reaches at most around 5.5 per
cent of the direct flux at its maximum.

This plot also shows the ratio of the diffuse to direct field in the
OTS and Ritzerveld approximations.  It is clear that the OTS
approximation predicts a reasonable average value for this ratio, but
overestimates the relative importance of the diffuse field in the core
of the nebula and underestimates it at the edge of the nebula.  In
contrast, Ritzerveld's formalism underestimates relative importance of
the diffuse field in the core but overestimates it at the edge of the
nebula, particularly for the case of equal absorption coefficients and
steeply decreasing density distibutions (as we saw in
Figure~\ref{f:radiat}, this is primarily due to underestimating the
direct field here).  In both cases, the first approximation would be
that the lateral diffuse field was one quarter the radiation
intensity, which would be a significant overestimate for the case of
equal absorption coefficients.

In Figure~\ref{f:shell}, we show the same results as previously for a
thin shell of uniform density between $0.9r_{\rm S}$ and $r_{\rm S}$.
In this case, Ritzerveld's approximation obviously underestimates the
diffuse field dramatically in the empty core of the nebula, as there
are no recombinations in this region, while the OTS approximation
overestimates the diffuse flux.  The radiation field is symmetric in
angle at all our sample points, which are at radii less than the inner
edge of the shell: this is to be expected, as the angle to the surface
from oppositely directed beams is the same and the surface brightness
is constant.  Particularly in the case with higher direct absorption
coefficient, limb-brightening is seen for the sample point at the
inner surface of the shell, corresponding to the higher direct flux
(and hence source function) at the $\tau_{\rm dif}=1$ surface in this
direction.

\subsection{Transfer using the diffusion approximation}

\begin{figure*}
\begin{tabular}{cc}
\includegraphics[width=6cm,angle=270]{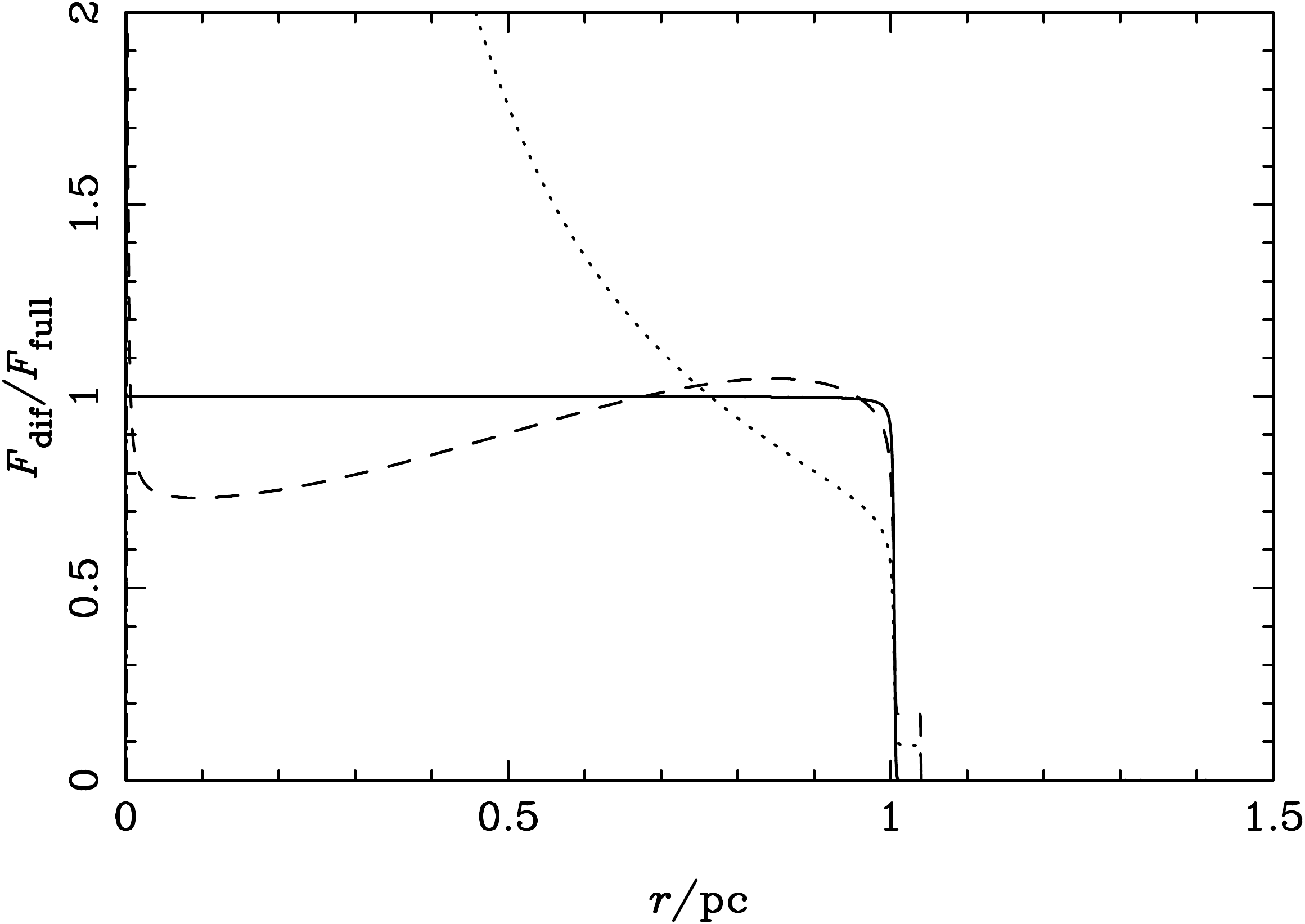} &
\includegraphics[width=6cm,angle=270]{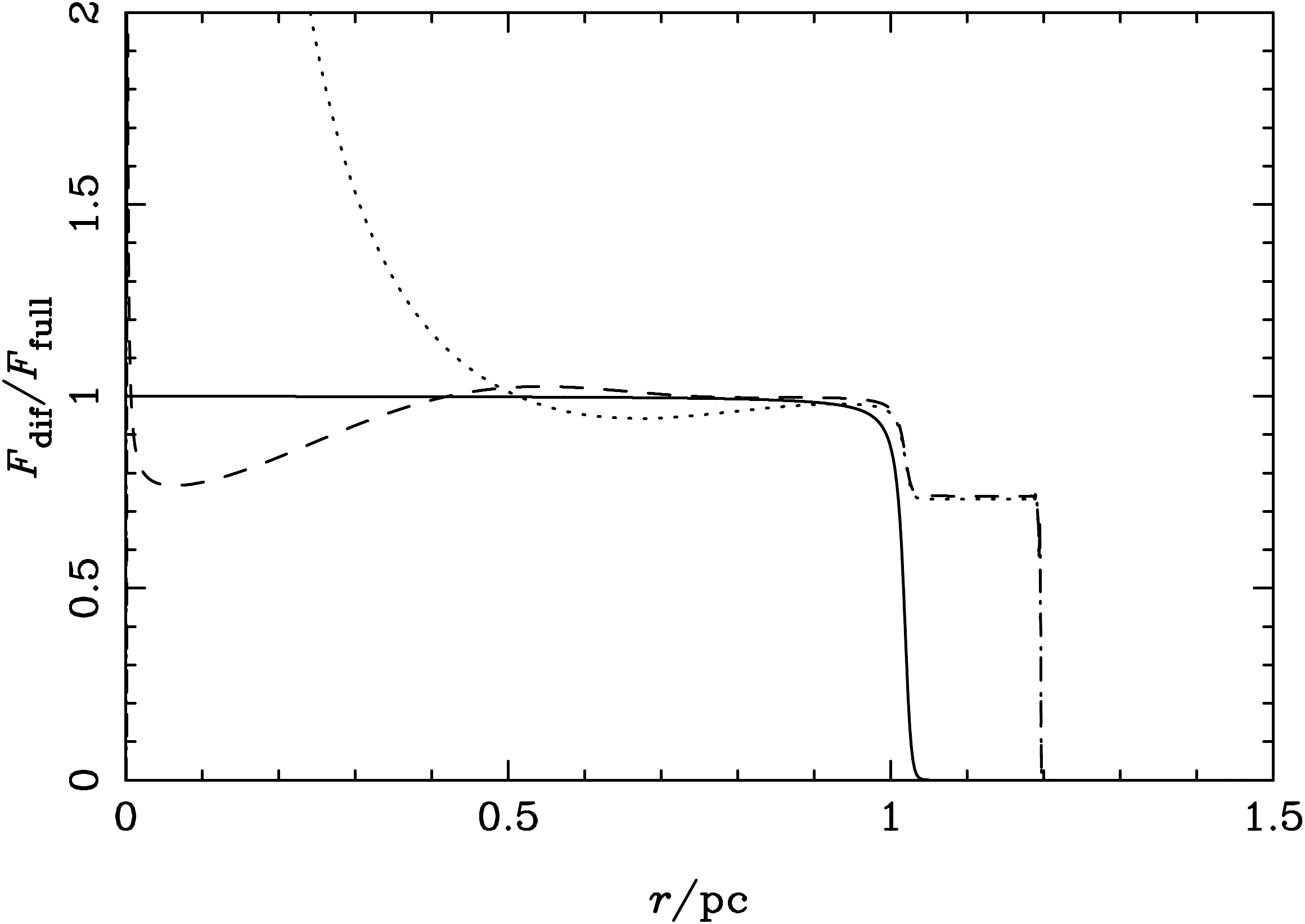} \\
\end{tabular}
\caption{Accuracy of diffuse field determination for uniform density
field.  Left column, with equal absorption coefficients, right column
with diffuse absorption coefficient 6 times that for direct photons.
Solid line -- ionization fraction; dotted line -- OTS diffuse field
divided by detailed transport diffuse field; dashed line -- diffusion
approximation diffuse field divided by detailed transport diffuse
field.}
\label{f:accuracy}
\end{figure*}

In Figure~\ref{f:accuracy}, we compare the accuracy of a diffusion
approximation solution to the diffuse field intensity to that of the
on-the-spot approximation.  The diffusion results are almost
everywhere more accurate than the OTS approximation, being in error by
less than 30 per cent even in the most difficult, central region of
the flow.  Both schemes are accurate at the edge of the nebula, where
the most interesting dynamical structures occur, although the
diffusion results are still significantly more precise.  

A diffusion approximation solver could be embedded in a dynamical code
by updating the ionization solution consistently with the sources
derived from the diffusion solver and iterating, in a similar fashion
to that which we use for the full field transfer.  Experimenting with
updating the ionization structure after both inward and outward
transfer are complete suggests that the solution converges more slowly
than if the ionization structure is updated on each sweep.  However,
in the context of a dynamical scheme, the initial approximation for
the ionization structure will have come from the previous timestep,
and hence would be expected to not be far from the correct solution.

The diffusion approximation allows processes such as the lateral
illumination of the tails of photoionized globules to be modelled
\citep{canto98,pavla01}.  However, the lateral illumination of
shadowed tails may be overstimated in cases where the beaming of the
diffuse field is expected to be strong \citep{melle06}: Eddington
tensor approaches \citep[e.g.\@][]{gonza05} may improve the results in
such limits.

\section{Discussion}

\subsection{Accuracy of the OTS approximation}

\begin{figure}
\begin{center}
\includegraphics[width=6cm,angle=270]{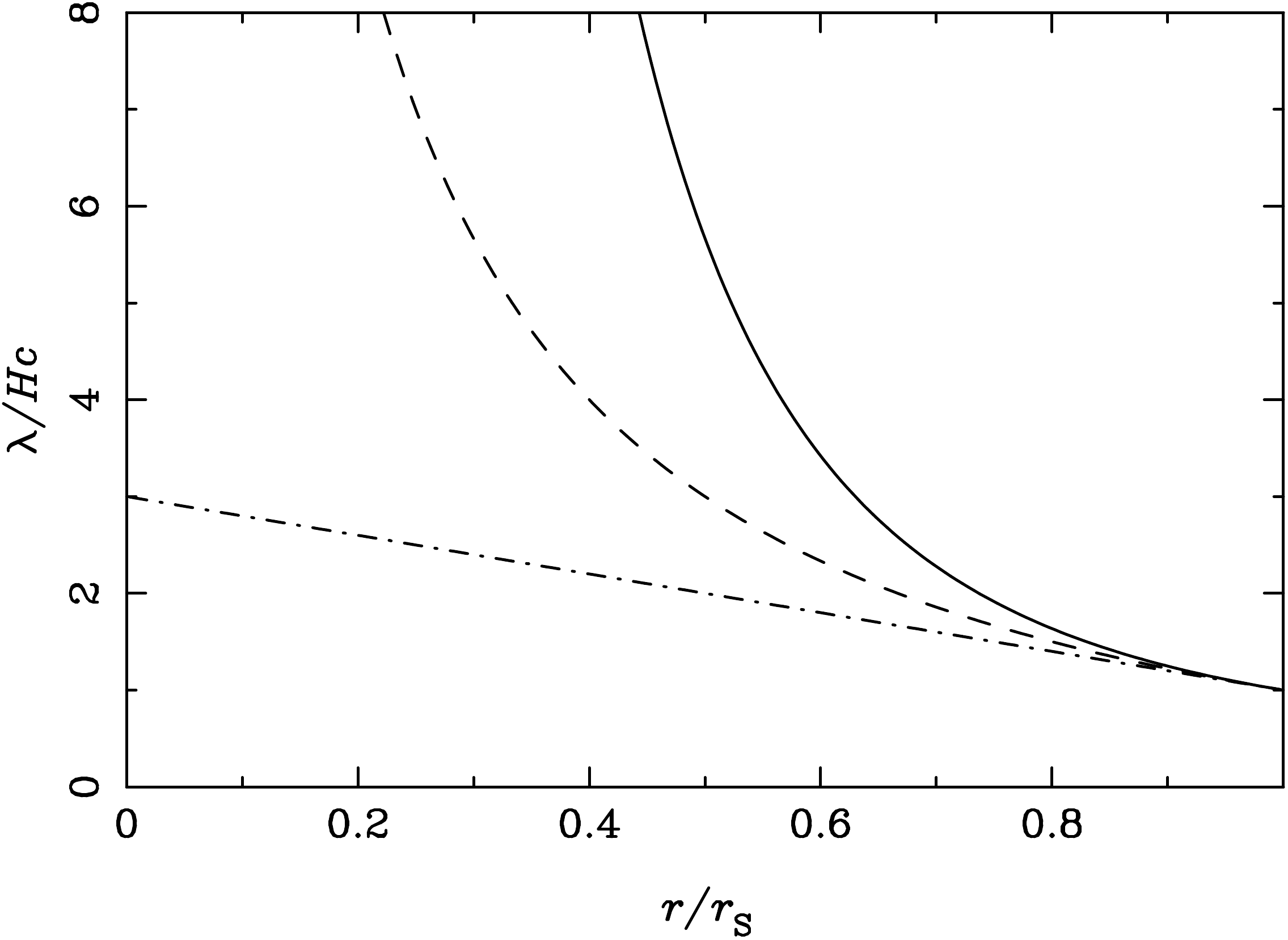}
\end{center}
\caption{Functional dependence from equation~(\protect\ref{e:otsacc})
for power-law density indexes $m=0$ (solid), $m=1$ (dashed), $m=2$
(dot-dashed).  This function shows where the OTS approximation for the
diffuse field is expected to be accurate (roughly for radii where the
ordinate is less than $a_1/a_0$).}
\label{f:otsacc}
\end{figure}
The accuracy of the OTS approximation may be evaluated by comparing
the scale of variation of the diffuse field source function, $H$, to
the mean free path for diffuse photons, $\lambda = 1/n(1-x)a_1$.

Using equation~(\ref{e:sdif}) for the source function, and the OTS
approximation equation~(\ref{e:casebbalance}) for the ionization
balance
\begin{equation}
S_{\rm dif} \simeq {1\over 4\pi} {\alpha_1 a_0\over \alpha_{\rm B} a_1} 
{L_{\rm dir}\over 4\pi r^2}.
\end{equation}
Note that we can still use the OTS approximation to close our
equations at this higher order.  The source function varies radially,
so that
\begin{eqnarray}
H^{-1} &=& \left\vert{1\over S_{\rm dif}} {dS_{\rm dif}\over dr}\right\vert  \\
& \simeq & \left\vert{1\over L_{\rm dir}} {dL_{\rm dir}\over dr} - 
{2\over r}\right\vert\\
& \simeq & {4\pi r^2\alpha_{\rm B} n^2 x^2\over L_{\rm dir}} + {2\over r},
\end{eqnarray}
where we have used equation~(\ref{e:ldirots}).  Now through most of
the H{\sc\,ii} region, $x\sim 1$, and hence
\begin{equation}
L_{\rm dir} \simeq  {4\pi r^2\alpha_{\rm B} n^2x^2\over n(1-x)a_0} 
= {a_1\over a_0} 4\pi r^2\alpha_{\rm B} n^2\lambda,
\end{equation}
and so
\begin{eqnarray}
{\lambda\over H} &\simeq& {a_0\over a_1} 
\left( 1+ { L_{\rm dir}\over 2\pi r^3 \alpha_{\rm B} n^2}\right) \\
&\simeq & {a_0\over a_1} 
\left(1+ {2\over n^2r^3}\int_r^{r_{\rm S}} n^2 r^2 dr\right),
\end{eqnarray}
where the integral is the result of equation~(\ref{e:ldirots}).  For a
density distribution $n\propto r^{-m}$ outside some specific radius,
\begin{equation}
{\lambda\over H} \simeq {a_0\over a_1} {1\over 3-2m}
\left(2(r/r_{\rm S})^{2m-3}+1-2m\right)\label{e:otsacc}
\end{equation}
(so long as $m\ne 3/2$).  In Figure~\ref{f:otsacc} we show the form of
the r.h.s.\@ of equation~(\ref{e:otsacc}) for $a_0=a_1$.  This
suggests that OTS approximation will be most accurate close to the
outer edge of H{\sc\,ii} regions.  When $a_0=a_1$, the OTS
approximation for the diffuse field is only marginally valid at any
point in the nebula.  For $a_0<a_1$, the OTS approximation may be
valid for a larger region at the outside of the nebula.  This is in
accord with the results we have already seen in Figure~\ref{f:radang};
equation~(\ref{e:otsacc}) is a simple formula which allows the
accuracy to be assessed for a variety of H{\sc\,ii} region structures.

In conclusion, while the OTS approximation gives accurate results for
the ionization structure in the H{\sc\,ii} region, as we have seen in
Figure~\ref{f:ionize}, the form of the diffuse field opacity
$n(1-x)a_1$ emphasizes differences in the interior of the region,
where $x\sim 1$.  Hence, as demonstrated for certain H{\sc\,ii} region
structures in Figure~\ref{f:radang} and confirmed in general by the
analysis in this section, the OTS approximation will give poor results
for the diffuse radiation field, particularly when $a_0\sim a_1$.

\subsection{Dust effects on the diffuse field}

As has been mentioned, there is some observational evidence for dust
absorption in well-observed H{\sc\,ii} regions.  The effect of such
dust absorption on nebular structure and emission has been widely
studied, recently by \cite{arthu04,dopit06}.

Dust absorption will act to further limit the importance of the
diffuse field.  This is true independent of the relative dust
absorption cross section at different frequencies, since the
absorption of direct photons reduces the net source of diffuse
photons.

To obtain a more accurate estimate of the effects of dust, we assume a
dust cross-section in EUV of $\sigma_{\rm g}\simeq 1.2\times
10^{-21}{\rm\,cm^2}$ per nucleon, with albedo $\omega\simeq0.5$
\citep{berto96}.  The dust scattering function is primarily
forward-beamed, so in the present analysis we will ignore the
dust-scattered continuum and take the effective absorption
cross-section to be $\sigma_{\rm g}' = (1-\omega)\sigma_{\rm g}$.

If we consider the OTS approximation, then for the diffuse field
\begin{equation}
J_{\rm dif} = 4\pi S = {\alpha_1 n x^2\over (1-x)a_1 + \sigma_{\rm
g}'}.
\label{e:dustots}
\end{equation}
We also have photoionization equilibrium, for which
equation~(\ref{e:ion}) still holds.  Substituting the OTS
approximation~(\ref{e:dustots}) into equation~(\ref{e:ion}), we find
\begin{equation}
J_{\rm dif} = 
{\alpha_1 (J_{\rm dir} a_0 + J_{\rm dif} a_1) \over
 \alpha_{\rm A} a_1 [1 + b(x)]},
\end{equation}
where $b(x) = \sigma_{\rm g}'/(1-x) a_1$.  Rearranging, using
$\alpha_{\rm B} = \alpha_{\rm A} - \alpha_1$, gives
\begin{equation}
{J_{\rm dif}\over J_{\rm dir}} = 
{a_0\over a_1}{\alpha_1 \over \alpha_{\rm B} + b(x) \alpha_{\rm A}}.
\end{equation}
Since in the presence of dust $b(x)>0$, the dust absorption will
always act to reduce the diffuse field, and will be significant effect
only where $b\sim 1$, i.e.\@ $(1-x) < \sigma_{\rm g}'/a_1 = 10^{-4}$.
For a uniform nebula in the case B approximation,
\begin{equation}
(1-x) \simeq {3\over a_0 n r_{\rm S}}{r^2\over r_{\rm S}^2}, 
\end{equation}
so, if there is a sufficient dust optical depth, the effects of dust
will be most significant close to the core of an H{\sc\,ii} region.

\section{Conclusions}

We present detailed calculations of radiation transfer within a
photoionized nebula, for a simplified physical model.  We find that
the diffuse field is indeed enhanced in the outer parts of such a
nebula if there are steep density gradients in the H{\sc\,ii} region,
particularly if the absorption cross section for diffuse and direct
photons is comparable.  These circumstances are not likely to be
typical.  We also find that when the diffuse radiation is relatively
strong, it is also strongly beamed in the radial direction, and so the
dynamical effect of the radiation field will in any case be similar to
the direct illumination.

We discuss the possibility of using a diffusion approximation for the
diffuse radiation field in two- and three-dimensional radiation
hydrodynamic calculations of H{\sc\,ii} regions.  This may allow the
diffuse field effects to be calculated to reasonable accuracy without
requiring a full radiation transfer calculation.

This paper has not considered in detail a number of additional
processes which may effect the radiation field within the nebulae,
such as the spectral hardening of radiation close to the ionization
front, dust absorption and scattering, and the effects of helium and
other heavy elements.  This is left for future work.

\section*{Acknowledgements}

WJH is grateful for financial support from DGAPA-UNAM, Mexico (PAPIIT
IN112006, IN110108 and IN100309).

\bsp

\label{lastpage}


\begin{thebibliography}{}

\bibitem[Abel et al.(1999)]{abele99}
Abel, T., Norman, M. L., Madau, P., 1999.  ApJ, 523, 66

\bibitem[Arthur et al.(2004)]{arthu04}
Arthur, S. J., Kurtz, S. E., Franco, J., Albarr\'{a}n, M. Y., 2004.
ApJ, 608, 282

\bibitem[Bertoldi \& Draine(1996)]{berto96}
Bertoldi, F., Draine, B. T., 1996.  ApJ, 458, 222

\bibitem[Cant\'o et al.(1998)]{canto98}
Cant\'o, J., Raga, A. C., Steffen, W., Shapiro, P., 1998.  ApJ 502,
695

\bibitem[Cesarsky et al.(2000)]{cesar00}
Cesarsky, D., Jones, A. P., Lequeux, J., Verstraete, L., 2000.  A\&A, 358, 708

\bibitem[Dopita et al.(2006)]{dopit06}
Dopita, M. A., et al., 2006.  ApJ, 639, 788

\bibitem[Franco et al.(1990)]{franc90} 
Franco, J., Tenorio-Tagle, G., Bodenheimer, P., 1990.  ApJ, 349, 126

\bibitem[Gonz\'alez \& Audit(2005)]{gonza05}
Gonz\'alez, M., Audit, E., 2005.  Ap\&SS, 298, 357

\bibitem[Hummer \& Seaton(1963)]{humme63}
Hummer, D. G., Seaton, M. J., 1963.  MNRAS, 125, 437

\bibitem[Mellema et al.(2006)]{melle06}
Mellema, G., Iliev, I., Alvarez, M., Shapiro, P., 2006.  New 
Astronomy, 11, 374.

\bibitem[Mihalas \& Weibel-Mihalas(1999)]{mihal99}
Mihalas, D., Weibel-Mihalas, B., 1999.  Foundations of Radiation 
Hydrodynamics, Dover: New York

\bibitem[O'Dell et al.(2007)]{odell07}
O'Dell, C. R., Henney, W. J., Ferland, G. J., 2007.  AJ, 133, 2343

\bibitem[Osterbrock and Ferland(2006)]{agn3}
Osterbrock, D. E., Ferland G.J., 2006.  
Astrophysics of Gaseous Nebulae and Active Galactic Nuclei, Second Edition,
University Science Books: Sausalito

\bibitem[Pavlakis et al.(2001)]{pavla01}
Pavlakis, K. G., Williams, R. J. R., Dyson, J. E., Falle, S. A. E. G.,
Hartquist, T. W., 2001.  A\&A 369, 263

\bibitem[Ritzerveld(2005)]{ritze05}
Ritzerveld, J., 2005.  A\&A 439, L23 (astro-ph/0506637)

\bibitem[Robberto et al.(2005)]{robbe05}
Robberto, M., et al., 2005.  AJ, 129, 1534

\bibitem[Rubin(1968)]{rubin68}
Rubin, R. H., 1968.  ApJ, 153, 761

\bibitem[Whalen \& Norman(2006)]{whale06}
Whalen, D. J., Norman, M. L., 2006.  ApJS, 162, 281


\bibitem[Williams(2002)]{willi02}
Williams, R.J.R., 2002.  MNRAS, 331, 693

\end{thebibliography}
\end{document}